\documentclass[twocolumn,longbibliography,aps,prc,superscriptaddress,showpacs,floatfix]{revtex4-1}
\usepackage{graphics,epsfig,graphicx}
\usepackage{comment}
\usepackage{amsmath}
\usepackage{bm}
\usepackage{bbm}
\usepackage{amsfonts}
\usepackage{cancel}
\usepackage{multirow}
\usepackage[dvipsnames]{xcolor}
\usepackage{natbib} 

\bibpunct{[}{]}{,}{n}{}{}

\newcommand{\pp}{pp}
\newcommand{\ppp}{ppp}
\newcommand{\pL}{p\Lambda}
\newcommand{\ppL}{pp\Lambda}
\newcommand{\nn}{nn}
\newcommand{\nnn}{nnn}
\newcommand{\nnL}{nn\Lambda}
\newcommand{\nL}{n\Lambda}

\begin{document}

\title{The $\pL$ and $\ppL$ correlation functions}

\author{E. Garrido}
\affiliation{Instituto de Estructura de la Materia, CSIC, Serrano 123, E-28006 Madrid, Spain}
\author{A. Kievsky}
\affiliation{Istituto Nazionale di Fisica Nucleare, Largo Pontecorvo 3, 56127 Pisa, Italy}
\author{M. Gattobigio}
\affiliation{Universit\'e C\^ote d'Azur, CNRS, Institut  de  Physique  de  Nice,
17 rue Julien Laupr\^etre, 06200 Nice, France}
\author{M. Viviani}
\affiliation{Istituto Nazionale di Fisica Nucleare, Largo Pontecorvo 3, 56127 Pisa, Italy}
\author{L.E.~Marcucci}
\affiliation{Physics Department, University of Pisa, Largo Pontecorvo 3, 56127 Pisa, Italy}
\affiliation{Istituto Nazionale di Fisica Nucleare, Largo Pontecorvo 3, 56127 Pisa, Italy}
\author{R. Del Grande}
\affiliation{Physik Department E62, Technische Universit\"at M\"unchen, James-Franck-Straße 1, 85748 Garching, Germany}
\author{L. Fabbietti}
\affiliation{Physik Department E62, Technische Universit\"at M\"unchen, James-Franck-Straße 1, 85748  Garching, Germany}
\author{D. Melnichenko}
\affiliation{Physik Department E62, Technische Universit\"at M\"unchen, James-Franck-Straße 1, 85748 Garching, Germany}

\begin{abstract}
In this work we present the study of $\pL$ and $\ppL$ scattering processes using femtoscopic correlation functions. This observable has been recently used to access the low-energy interaction of hadrons emitted in the final state of high-energy collisions, delivering unprecedented precision information of the interaction among strange hadrons. The formalism for particle pairs is well established and it relates the measured correlation functions with the scattering wave function and the emission source. 
In the present work we analyze the $NN\Lambda$ scattering in free space and relate the corresponding wave function to the $\ppL$ correlation measurement performed by the ALICE collaboration. The three-body problem is solved using the hyperspherical adiabatic basis. Regarding the $\pL$ and $\ppL$ interactions, different models are used and their impact on the correlation function is studied. The three body force considered in this work is anchored to describe the binding energy of the hypertriton and to give a good description of the two four-body hypernuclei. As a main result we have observed a huge, low-energy peak in the $\ppL$ correlation function, mainly produced by the $J^\pi=1/2^+$ three-body state. The study of this peak from an experimental as well as a theoretical point of view will provide important constraints to the two- and three-body interactions. 
\end{abstract}


\maketitle

\section{Introduction}

The study of the strong interaction between nucleons ($N$) and $\Lambda$ hyperons has a long tradition, dating back to the discovery of the first $\Lambda$-hypernucleus \cite{danysz1953delayed}. 
Realistic interactions were constructed using meson-exchange models, see Refs.~\cite{lagaris,bodmer1,bodmer}, and used to describe the $\Lambda$ separation energy in hypernuclei. In these studies the prominent role of the
$NN\Lambda$ three-body interaction was immediately understood. In fact it was not possible to describe the $\Lambda$ separation energy considering only two-body forces.
This triggered an increasing interest in the dynamics of those systems and, along the years, many new data became available from $\pL$ scattering experiments~\cite{PhysRev.175.1735,PhysRev.173.1452,PhysRevLett.127.272303} and measurements of the binding energies of hypernuclei~\cite{HASHIMOTO2006564,RevModPhys.88.035004,ALICE2019134905,STAR}. In particular, detailed studies of the $NN\Lambda$ interactions appeared in order to understand their possible impact in the description of compact systems~\cite{Tolos:2020aln,pederiva}, having baryon densities far beyond the nuclear saturation one. In recent years the $N\Lambda$ and $NN\Lambda$ interactions have been studied using the effective field theory (EFT) framework as it allows to systematically improve the precision of the calculations
by considering higher-order terms in the Lagrangian (for a recent review see Ref.~\cite{pets2020}). 
All these improvements renewed the interest in having accurate data for $N\Lambda$ and $NN\Lambda$ systems as they are necessary to determine the, a priori unknown, low-energy constants (LECs) associated with the contact
interactions in the theory. A precise determination of the scattering parameters requires measuring the scattering of hadrons at low relative momenta, down to the energy threshold. Conventional scattering experiments face several difficulties, as the realization of low-energy stable beams with strange hadrons is challenging. Moreover, data on the low-energy scattering of a $\Lambda$ hyperon with few-nucleons do not exist yet. 

Recently, the femtoscopy technique \cite{Wiedemann:1999qn,Heinz,fem2} has been used at the Large Hadron Collider (LHC)
to perform new high-precision studies of the low-energy interactions between hadrons. This experimental method exploits the production and emission of hadrons at relative distances of the order of a femtometer in $pp$ and $p$-nucleus collisions,
to study their final state interaction~\cite{lfabb2021}.
The interaction between hadrons appears as a correlation signal in the momentum distributions
of the detected particles which can be measured in the form of a correlation function. This function
depends on the emission process, which is the source of hadrons, as well as on the final state interaction of the emitted particles.
By measuring correlated particle pairs or triplets at low relative energies and comparing the 
results of the measurements to theoretical predictions, it is possible to extract information
on the two-body hadron-hadron interaction and, eventually, on the three-body interaction.

The success of this method in the two-body sector has been demonstrated by the ALICE Collaboration.
Several high-precision measurements involving strange hadrons have been performed, 
 making it possible to test lattice calculations and challenge EFT results
(for a recent review see Ref.~\cite{lfabb2021} and references therein). In particular, the most precise data on the $\pL$ system has been delivered in Ref.~\cite{pLambda}. In this work the pair correlation function has been measured and information on the $p\Lambda$ interaction in the low energy region down to the threshold, at relative momenta $k < 45$ MeV/$c$, could be accessed for the first time. 
The next challenge is to achieve the same precision for three-particle correlations and to test the hadronic interactions in three-body systems. Promising experimental results for three baryons, as
the $\ppp$, $\ppL$ and $pd$ correlation functions, have been already obtained in Refs.~\cite{femtoppp, femtopd} from the analysis of the data acquired during the LHC Run 2 period and significant improvements in accuracy are expected from the ongoing Run 3 data taking \cite{ALICE:2020fuk}. 

The theoretical interpretation of such measurements requires to calculate the correlation function as a convolution between a source function, which models the emission of hadrons, and the norm of the scattering wave function. The latter incorporates information
on the interaction between the particles and, at low values of the relative momenta, it can be computed by solving
the Schr\"odinger equation~\cite{fempp}. The three-nucleon $\ppp$ and $pd$ correlation functions have been subject of recent theoretical studies~\cite{pppth,pdtheory}. As the interaction of nucleons is already known with a high level of accuracy, these studies allowed to establish the three-body correlation function formalism. In particular, those studies have shown that a full solution of the three-body problem is needed to reproduce the complex structure
of the measured correlation signal. From a partial wave expansion of the scattering wave function, it has been shown the importance
of the different terms in the description of the observable. 

The experimental study of the correlation function of three nucleons can deliver complementary information with respect to traditional experiments in nuclear physics.
 For example the scattering of three protons cannot be carried out in nuclear laboratories since experiments having three in-going nucleons are very challenging. Similarly, the $\ppL$ correlation function measurement represents  a unique doorway to access the low-energy $NN\Lambda$ dynamics and to extract new information about the $N\Lambda$ and $NN\Lambda$ interactions.

In the present study we will describe theoretically the $\pL$ and the 
$\ppL$ correlation functions and compare them to the available experimental data.
For the $\pL$ system two main ingredients have to be considered: the knowledge of the source function and the $\pL$ interaction needed to calculate the scattering wave function at different energies. Following previous studies, the former is modeled as the product of single Gaussian emitters, depending on the size of the source \cite{ALICEsource,ALICEsourceErr}. 

To compute the $\ppL$ correlation function the $\ppL$ scattering wave function has to be calculated. This is a challenging problem in which the description of the three-body continuum is needed considering that two of the three particles are charged. Following the studies we have carried out to describe the $pd$ and $ppp$ correlation functions~\cite{pdtheory,pppth,pppC}, we use the hyperspherical adiabatic (HA) basis~\cite{garrido} to study the $\nnL$ and $\ppL$ continuum states. 
The investigation of these two systems will allow us to observe the differences produced by the Coulomb force. Moreover, in this analysis, we make a partial wave decomposition of the scattering wave function in terms of states having well defined values of total angular momentum and parity, $J^\pi$. In this way the relevance of the different expansion terms can be assessed. As a main result of this analysis we have observed a huge peak, produced by the $J^\pi=1/2^+$ state, dominating the correlation function at low energies. The sensitivity of the peak with respect to the $\pL$ and $\ppL$ interactions, as well as with respect to the
source size, is an important part of the present study as a tool to extract information about the unknown three body interaction.

The paper is organized as follows: In Section~\ref{sect2} we investigate the $\pL$ system, giving the details of the $\pL$ potentials used in this work and comparing the computed correlation function with the available experimental data. The formal details of the three-particle correlation function are given in Section~\ref{sect3}, were we describe as well the three-body force used in the
calculations. This formalism is employed in Sections~\ref{sect4} and \ref{sect5}
to investigate the $\nnL$ and $\ppL$ correlation functions, respectively. The comparison with the experimental data is shown in Section~\ref{sect6}, and we finish in Section~\ref{sect7} with the summary and the conclusions.

\section{The $\pL$ correlation function}
\label{sect2}

The study of the $\pL$ scattering process with femtoscopic data requires knowledge of the emission source functions $S_i (\bm{r}_i)$ for the proton in the position $\bm{r}_1$ and for the $\Lambda$ hyperon in the position $\bm{r}_2$, as required in the definition of the correlation function~\cite{monrow}
 \begin{equation}
	 C_{\pL}(k)=\int d\bm{r}_1d\bm{r}_2 \, S_1(\bm{r}_1) S_2(\bm{r}_2)
	 |\Psi(\bm{r}_1,\bm{r}_2)|^2,
 \label{cor1e2}
\end{equation}
where $\Psi(\bm{r}_1,\bm{r}_2)$ is the two-body scattering wave function.

In the present work, the single particle source functions are chosen to have the same Gaussian form used in previous studies, i.e.,
 \begin{equation}
	 S_i(\bm{r})=\frac{1}{(2\pi R_{M,i}^2)^{3/2}}e^{-r_i^2/2R_{M,i}^2} \ ,
 \label{spf12}
\end{equation}
where $R_{M,i}$ denotes the single particle source radius for the hadron specie $i$. 

Introducing the center of mass coordinate and momentum, $\bm{R}$ and $\bm{P}$, and the relative coordinate and momentum, $\bm{r}$ and $\bm{k}$, we have that  
$\Psi(\bm{r}_1,\bm{r}_2)=e^{i\bm{R}\cdot \bm{P}}\Psi_k(\bm{r})$, and the
correlation function is then given in the pair rest frame by the Koonin-Pratt equation~\cite{Koonin,Pratt}
 \begin{equation}
	 C_{\pL}(k)=\int d\bm{r} \,  S_{12}(\bm{r}) |\Psi_k(\bm{r})|^2,
 \label{corr12}
\end{equation}
where 
 \begin{equation}
	 S_{12}(\bm{r})=\int d\bm{R}\ S_1(\bm{r}_1)\ S_2(\bm{r}_2)=
	 \frac{1}{(4\pi r_{0}^2)^{3/2}}e^{-r^2/4r_{0}^2} \, ,
 \label{tpf12}
\end{equation}
is the pair source function which depends on the pair relative distance $r$ and where
\begin{equation}
    r_{0} = \frac{1}{\sqrt{2}} \sqrt{R_{M,1}^2 + R_{M,2}^2}
    \label{eq:r0}
\end{equation}
is the two-particle source radius. 

The source function in Eq.~(\ref{tpf12}) is normalized to unity in coordinate space, and it can therefore be interpreted as the probability to emit the two particles at a relative distance $r$. 
In order to make comparisons with the $\pL$ measurement performed by ALICE~\cite{pLambda},
the source radius $r_0$ is fixed according to the source model developed in Ref.~\cite{ALICEsource} for $\pp$ collisions at the LHC. The details are discussed in Section~\ref{sec:Results2B}.

The term $ |\Psi_k|^2$ in Eq.~\eqref{corr12} is the square of the scattering wave function of the two particles at energy 
$E=\hbar^2k^2/2\mu$, being $\mu$ the reduced mass, $1/\mu=1/m + 1/M$, with $m$
the nucleon mass and $M$ the mass of the $\Lambda$ particle. 

As a preliminary step to analyze  the scattering process,
we discuss the scattering wave function considering no interaction between the particles. The
two-body wave function is therefore given just by the plane wave
\begin{equation}
\Psi_s^0=\frac{1}{\sqrt{N_S}}e^{i\bm{k}\cdot\bm{r}} \sum_{SS_z} \chi_{SS_z},
\label{freewf}
\end{equation}
where $\chi_{SS_z}$ is the spin function arising from the coupling of two spin-$\frac{1}{2}$ particles
to $S=0,1$, with projection $S_z$, and $N_S=\displaystyle\sum_{SS_z} 1 =\sum_S(2S+1)=4$ is the number of spin states.
Considering the usual partial wave expansion of the plane wave, Eq.~(\ref{freewf}) can also be written as
\begin{eqnarray}
\lefteqn{
	\Psi^0_s=  } \label{eq:psi0} \\ & &
 \frac{4\pi}{\sqrt{N_S}}
         \sum_{JJ_z}\sum_{\ell m S S_z} i^\ell j_\ell(kr) (\ell m S S_z|JJ_z)
	 {\cal Y}^{JJ_z}_{\ell S}(\Omega_r) Y^*_{\ell m}(\Omega_k),
  \nonumber
\end{eqnarray}
where $\ell$, $S$, and $J$ are the relative orbital angular momentum, total spin, and total
angular momentum, respectively, with projections $m$, $S_z$, and $J_z$, and where $\Omega_r$ and $\Omega_k$
are the polar and azimuthal angles describing the directions of the relative coordinate ($\bm{r}$) 
and the relative momentum ($\bm{k}$). In the expression above, $j_\ell(kr)$ is the regular Bessel function 
and
\begin{equation}
{\cal Y}_{\ell S}^{JJ_z}(\Omega_r)=\sum_{m S_z} (\ell m \, S S_z|JJ_z) Y_{\ell m}(\Omega_r) \chi_{SS_z} \,,
 \label{coup0}
\end{equation}
where ${Y}_{\ell m}(\Omega_r)$ is a spherical harmonic function. 

The norm of the scattering wave function, $|\Psi^0_s|^2_\Omega$, is defined as the average over the angular 
coordinates of the square of the wave function, i.e., 
\begin{equation}
    |\Psi^0_s|^2_\Omega = \frac{1}{(4\pi)^2}\int d\Omega_r  \int d\Omega_k |\Psi^0_s|^2,
    \label{norm}
\end{equation}
which after introducing Eq.~(\ref{freewf}), becomes 
\begin{equation}
|\Psi^0_s|^2_\Omega=1 \ .
\label{normfac}
\end{equation}
Making use of Eq.~(\ref{eq:psi0}), the norm in Eq.~(\ref{norm}) can also be written as
\begin{equation}
|\Psi^0_s|^2_\Omega= \frac{1}{N_S} \sum_{\ell S} j^2_\ell(\eta,kr) 
(2\ell + 1) (2S+1).
\end{equation}
 Recalling that $\sum_\ell j^2_\ell(\eta,kr) (2\ell+1)=1$ and $N_S=\sum_S(2S+1)$ the result $|\Psi^0_s|^2=1$ is recovered.

When the two particles interact, it is convenient to write the scattering wave function in the form analogous to Eq.~(\ref{eq:psi0})
\begin{equation}
	\Psi_s=\frac{4\pi}{\sqrt{N_S}}\sum_{JJ_z}\sum_{\ell m S S_z} i^\ell \Psi_{\ell S}^{JJ_z} (\ell m S S_z|JJ_z) 
        Y^*_{\ell m}(\Omega_k),
 \label{eq:psi1}
\end{equation}
where $\Psi_{\ell S}^{JJ_z}$ is the coordinate wave
function of the system with quantum numbers $\ell$, $S$, $J$, and $J_z$, which reduces
to $\Psi_{\ell S}^{JJ_z}=j_\ell(kr) {\cal Y}^{JJ_z}_{\ell S}(\Omega_r)$ when the interaction is not present. 

More precisely, if the nuclear short-range $\pL$ interaction is considered, the coordinate wave function takes the form
\begin{equation}
    \Psi_{\ell S}^{JJ_z}= \sum_{\lambda S'} 
  \frac{u^{\lambda S'}_{\ell S}\!(k,r)}{kr} {\cal Y}^{JJ_z}_{\lambda S'}(\hat r),
  \label{radeq}
\end{equation}
where we assume that, given an incoming channel with orbital angular momentum and spin $\{\ell, S\}$, 
the short-range interaction can mix it
with an outgoing channel with quantum numbers $\{\lambda, S'\}$.

The general large distance behavior of the radial functions in Eq.~(\ref{radeq}) is given by
\begin{equation}
    u^{\lambda S}_{\ell S}\rightarrow \delta_{\lambda \ell} kr [j_\ell(kr)+T_{\lambda \ell}^S {\cal O}_\ell(kr)],
\label{eq14}
\end{equation}
where for simplicity we have assumed that the potential is diagonal in the spin channels, i.e., $S=S'$. Moreover, if the potential does not couple the different $\ell-$channels either, we have that $u_\ell^\lambda \equiv u_{\ell S}$, and 
${\cal O}_\ell(kr)=\eta_\ell(kr)+i j_\ell(kr)$ describes the outgoing wave function ($j_\ell$ and $\eta_\ell$ are the regular and irregular spherical Bessel functions). In Eq.~(\ref{eq14})  $T^S_{\lambda\ell}$ denotes the $T$-matrix elements.

Therefore, from Eqs.~(\ref{eq:psi1}) and (\ref{radeq}), and following the same procedure as in the free case, the 
norm of the wave function results
\begin{equation}
|\Psi_s|^2_\Omega= \frac{1}{N_S}
	\sum_{\ell S} \left(\frac{u_{\ell S}(k,r)}{kr}\right)^2 (2\ell +1) (2S+1) \ .
 \label{norm2}
\end{equation}

Finally, to compute the correlation function we will use the Gaussian source function
given in Eq.~(\ref{tpf12}), and we will also take into account that,  since the source is spherical, the correlation function can be computed as given in Eq.~(\ref{corr12}) but replacing $|\Psi_s|^2$ by $|\Psi_s|^2_\Omega$ as defined in Eq.(\ref{norm}) \cite{pppth}.  

For the specific case where the short-range interaction 
is limited to act on the $\ell=0$ state only, the norm given in Eq.~(\ref{norm2}) can be written as
\begin{equation}
	|\Psi_s|^2_\Omega= 1
	+\frac{1}{4} \left(\frac{u_{00}(k,r)}{kr}\right)^2
	+\frac{3}{4} \left(\frac{u_{01}(k,r)}{kr}\right)^2- j^2_0(kr),
\end{equation}
where we have added and subtracted the $\ell=0$ free case contribution in the spin channels $S=0,1$.
The correlation function results in
\begin{eqnarray}
\lefteqn{
	C_{\pL}(k)= 1 + }  \label{eq:cpl} \\ & & \hspace*{-5mm}
 \int d{\bm{r}}\, S_{12}(r)\left[\frac{1}{4} \left(\frac{u_{00}(k,r)}{kr}\right)^2
	+\frac{3}{4} \left(\frac{u_{01}(k,r)}{kr}\right)^2- j^2_0(kr)\right].
	\nonumber 
\end{eqnarray}

\subsection{Low-energy $\pL$ interaction}

The $s$-wave functions $u_{0S}(k,r)$ introduced above are calculated using the $\pL$
interaction. This interaction has been extensively discussed in the literature. A recent
review can be found in Ref.~\cite{pets2020}. In that reference, as well as in
Refs.~\cite{haid2023,NLO19},
the $s$-wave scattering lengths and effective ranges are given as a result
of the analysis of the available scattering data using a chiral-EFT framework at 
next-to leading order (potentials NLO13 and NLO19) and at next-to-next leading
order (potential SMS N2LO). The values of the low-energy
scattering parameters are collected 
in Table I for the different cutoffs, $C$, used in the analysis of the two possible spin $S=0,1$ channels.

\begin{table*}[t]
\begin{tabular}{l|cccccc|cccc|ccc}
	      &NLO13  &       &       &       &       &       & NLO19 &       & &       & SMS N2LO &   &     \cr
$ C $(MeV)    & 450   & 500   & 550   & 600   & 650   & 700   & 500   & 550   & 600   & 650   & 500   & 550   & 600  \cr
\hline
$a_0\,$(fm)   & -2.90 & -2.91 & -2.91 & -2.91 & -2.90 & -2.90 & -2.91 & -2.90 & -2.91 & -2.90 & -2.80 & -2.79 & -2.80  \cr
$r_e^0\,$(fm) &  2.64 &  2.86 &  2.84 &  2.78 &  2.65 &  2.56 &  3.10 &  2.93 &  2.78 &  2.65 &  2.82 &  2.89 &  2.68  \cr
\hline
$a_1\,$(fm)   & -1.70 & -1.61 & -1.52 & -1.54 & -1.51 & -1.48 & -1.52 & -1.46 & -1.41 & -1.40 & -1.56 & -1.58 & -1.56 \cr
$r_e^1\,$(fm) &  3.44 &  3.05 &  2.83 &  2.72 &  2.64 &  2.62 &  2.62 &  2.61 &  2.53 &  2.59 &  3.16 &  3.09 &  3.17  \cr
\end{tabular}
	\caption{The scattering lengths $a_S$ and effective ranges $r^S_e$, in spin channels $S=0,1$,
	predicted by the chiral-EFT potentials NLO13, NLO19 and SMS N2LO
	for the different values of the cutoff $C$.}
 \label{Tab1}
\end{table*}

In the following we introduce a low-energy description of the $\pL$ interaction
using a spin-dependent interaction of the Gaussian form

\begin{equation}
V_{\pL}(r)= \sum_S V_S e^{-(r/r_s)^2} {\cal P}_ {0,S} 
\label{eq:gauspot}
\end{equation}
where ${\cal P}_ {0,S}$ is a projector on the $\ell=0$ state of the $\pL$ system
and on the spin channel $S=0,1$. 

\begin{table*}[!ht]
\begin{tabular}{l|cccccc|cccc|ccc}
            & NLO13 &       &       &       &       &       &NLO19  &       &       &       & SMS N2LO &    &    \cr
$ C $(MeV)  & 450   & 500   & 550   & 600   & 650   & 700   & 500   & 550   & 600   & 650   & 500   & 550   & 600      \cr
\hline
$V_0\,$(MeV)&-35.130&-30.180&-30.574&-31.851&-34.831&-37.198&-25.954&-28.817&-31.851&-34.831&-31.140&-29.753&-34.273   \cr
$r_0\,$(fm) & 1.375 & 1.467 & 1.459 & 1.434 & 1.380 & 1.342 & 1.563 & 1.495 & 1.434 & 1.380 & 1.439 & 1.466 & 1.382  \cr
\hline
$V_1\,$(MeV)&-23.239&-29.205&-33.839&-36.258&-38.455&-39.143&-38.984&-39.470&-42.055&-40.373&-27.544&-28.609&-27.392 \cr
$r_1\,$(fm) & 1.482 & 1.338 & 1.247 & 1.216 & 1.183 & 1.170 & 1.178 & 1.163 & 1.126 & 1.143 & 1.361 & 1.344 & 1.364 \cr
\hline
$B(^3_\Lambda{\rm H})\,$(MeV)  & 2.949 & 2.873 & 2.879 & 2.925 & 2.985 & 3.027 &
2.792 & 2.839 & 2.904 & 3.255 & 2.819 & 2.799 & 2.878\\
\hline
$W_3\,$(MeV)  & 12.834 & 11.83 & 11.733 & 12.32 & 12.873 & 13.224 & 10.545& 11.056
	& 11.795 & 12.294 & 10.65 & 10.375 & 11.4\\
$\rho_3\,$(fm) & 2.0 & 2.0 & 2.0 & 2.0 & 2.0 & 2.0 & 2.0 & 2.0 & 2.0 & 2.0 & 2.0
& 2.0 & 2.0\\
\hline

\end{tabular}
	\caption{ The strength $V_S$ and range $r_S$ of the Gaussian potentials
	in the two spin channels $S=0,1$, reproducing the low-energy parameters given 
	in Table~\ref{Tab1}. The row with $B(^3_\Lambda{\rm H})$ shows the binding energy of the
	hypernucleus $^3_\Lambda$H. The last two rows give the strength and range of the	three-body force needed to describe the experimental value of
	$B(^3_\Lambda{\rm H})$.}
 \label{Tab2}
\end{table*}

The coupling strength and range parameters, $V_S$ and $r_s$ respectively,
are determined by reproducing the scattering length and effective range values
of the chiral potentials given in Table~\ref{Tab1}. The specific values are quoted in Table~\ref{Tab2}.
In this way we construct 
different sets of the Gaussian potential and we study the impact of the different 
parameterizations on the correlation function. The low-energy Gaussian representation of the $\pL$ interaction is justified in the context
of a leading order effective interaction based on contact interactions, see for example Ref.~\cite{contessi2018}. 

In Table~\ref{Tab2} the binding energy of the hypertriton, $B(^3_\Lambda{\rm H})$, predicted by these Gaussian potentials is also shown. The last two rows give the strength and range of a repulsive three-body interaction introduced to reproduce the experimental values of the hypertriton binding energy, $B(^3_\Lambda{\rm H})\approx 2.39\,$MeV. Details of the used three-body interaction are given in Section~\ref{sec:NNLforce}.
In addition to the Gaussian interactions, we consider as well the Bodmer-Usmani $\pL$ interaction~\cite{bodmer1,bodmer}
\begin{equation}
    V_{\pL}^{BU}=V_C(r)(1-\epsilon +\epsilon P_x)+0.25 \, V_G \, T^2_\pi(r)
    \, \mathbf{\sigma}_\Lambda\cdot\mathbf{\sigma}_p,
    \label{bodus}
\end{equation}
with $P_x$ the space-exchange operator and $\sigma_\Lambda$, $\sigma_p$ the Pauli matrices of the $\Lambda$ and proton respectively. The $r$-dependence of the potentials has the form $V_C(r)=W_C/(1+e^{(r-\bar r)/a}) -\bar v\, T^2_\pi(r)$ where $T_\pi(r)$ is the pion exchange tensor potential. \\
With the values of the
parameters $\epsilon=0$, $\bar r=0.5$~fm, $a=0.2$~fm, $\bar v=6.20$~MeV, $V_G=0.25$~MeV, $W_C=2137$~MeV, the Bodmer-Usmani potential predicts the following values
of the low energy scattering quantities: $a_0=-2.88$~fm, $r_e=2.87$~fm and 
$a_1=-1.66$~fm, $r_e=3.67$~fm and a binding energy of the hypertriton of
$B(^3_\Lambda{\rm H})=2.73$~MeV. These values are not very different to
those given in Table~\ref{Tab1} and Table~\ref{Tab2}, obtained using the chiral-EFT potentials and their Gaussian representations, respectively.

\subsection{Results and comparison to experimental data}\label{sec:Results2B}

In the following we show the results of the correlation function $C_{\pL}(k)$
given in Eq.~(\ref{eq:cpl}) using the different potential models to compute the
scattering wave functions $u_{00}(r)$ and $u_{01}(r)$, and we compare the calculations to the ALICE data by fixing the source radius $r_0$ and considering experimental corrections. 

The modeling of the source in $\pp$ collisions at
the LHC has been extensively studied in several recent
works~\cite{ALICEsource,ALICEsourceErr,Mihaylov:2023pyl,alicecollaboration2023common}. In Ref. \cite{ALICEsource} it was demonstrated that with the proper inclusion of the short-lived decaying resonances, the Gaussian core radius obtained from $\pL$ correlation measurements in different transverse mass $m_T$ intervals is in agreement with the $\pp$ results, 
supporting the existence of a common emitting source for all hadron pairs in $\pp$ collisions. 
The assumption of a common source for all hadrons was used to test the interaction models of 
several particle pairs and to access their low energy scattering properties through correlation function measurements~\cite{lfabb2021}. 

The source radius for the pairs of interest is determined from the transverse mass, $m_T$, scaling obtained from $\pp$ correlation measurements, considering the effective enlargement induced by the short-lived decaying 
resonances. The average transverse mass of the $\pL$ pairs in the ALICE data sample is $\langle m_T \rangle = 1.55$ GeV/$c^2$~\cite{pLambda}. This corresponds to an effective Gaussian radius of $r_0 = 1.15 \pm 0.04$ fm, which includes the effects induced by the short-lived resonances decaying into $p$ and/or $\Lambda$ in the final state~\cite{pLambda}. The uncertainty is propagated from the $\pp$ measurement including 10\% variations of the resonance production yields and fractions adopted in the source model. 

\begin{figure}[t]
\includegraphics[width=\columnwidth]{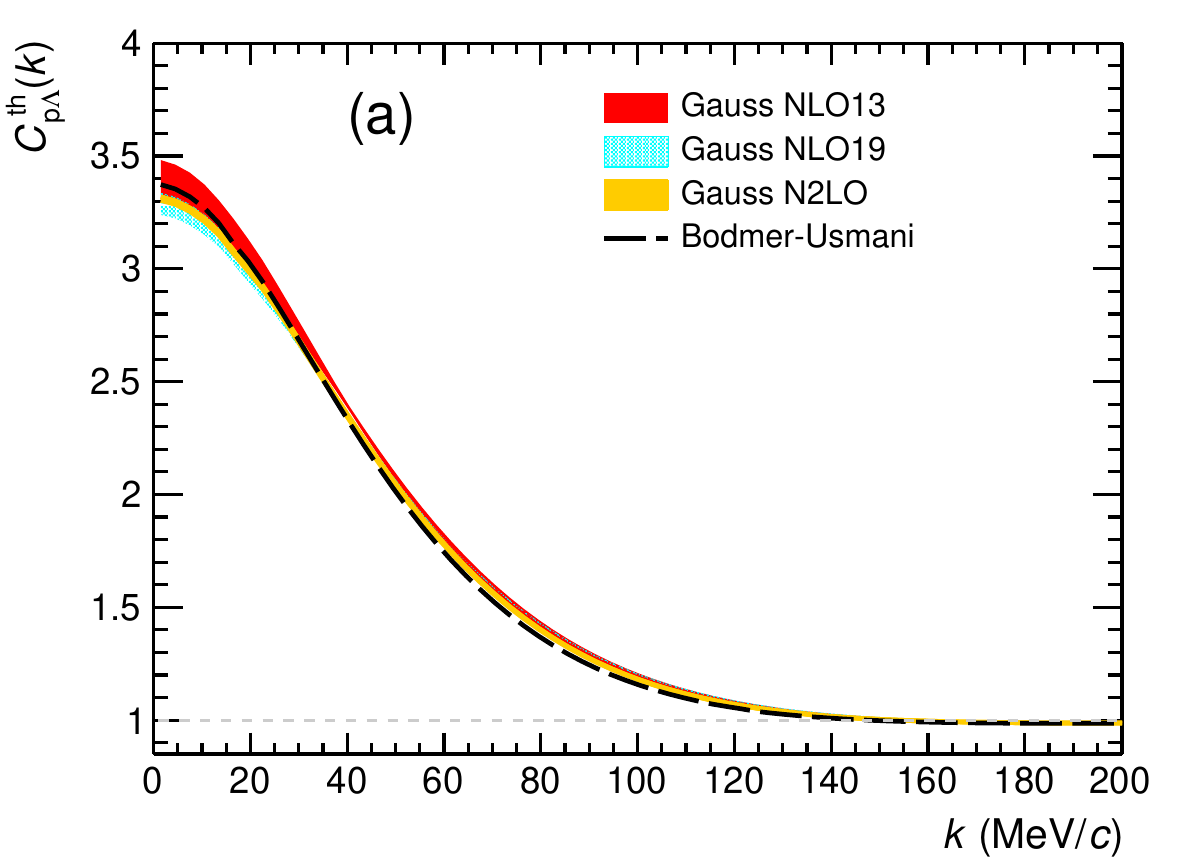}
 \includegraphics[width=\columnwidth]{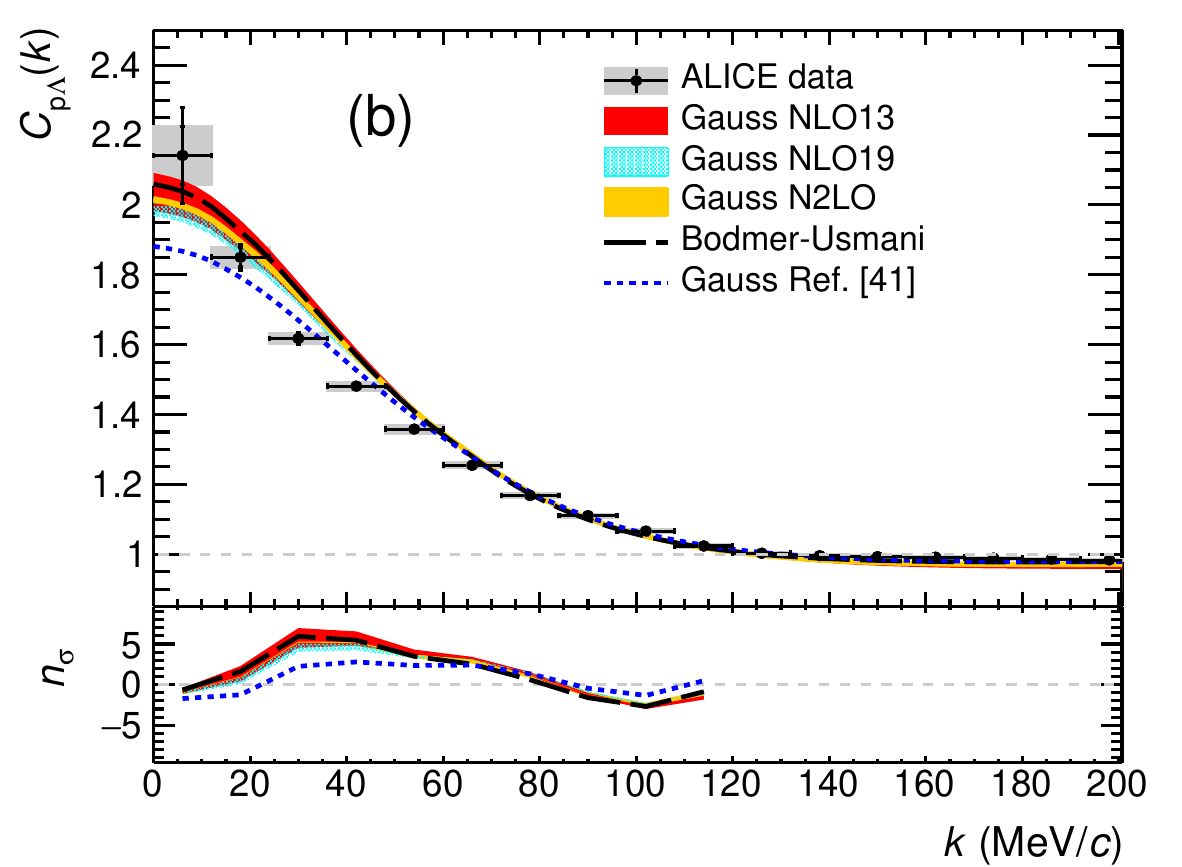}
	\caption{(a) $\pL$ correlation function calculated for the
	different potential models, see text for details. $C_{\pL}^\mathrm{th} (k)$ refers to the use of Eq.~(\ref{eq:cpl}). (b) Corrected correlation
	function as defined in Eq.~(\ref{eq:cplc}). In both panels, the source radius
	$r_0=1.19\,$fm is used to calculate the genuine $\pL$ correlation function.}
 \label{fig1}
\end{figure}

In Fig.~\ref{fig1} we show 
different bands for the Gaussian potential models reproducing the low energy parameters
at the different cutoff values. The Gaussian representation of the NLO13, NLO19
and N2LO potentials are collected in the red, cyan, and orange bands
respectively. Moreover, the result using the Bodmer-Usmani potential
is also given. Figure~\ref{fig1}a shows the direct calculation of the correlation function $C^\mathrm{th}_{\pL}(k)$, whereas Fig.~\ref{fig1}b shows the comparison to the ALICE measurement. The experimental correlation function additionally includes the contributions from residual (feed-down) correlations and misidentified particles.

The feed-down contributions  originate from the detection of proton or $\Lambda$ particles which are emitted in the decay of long-lived resonances. In the ALICE data, the dominant feed-down contributions arise from the decay of $\Sigma^0$, $\Xi^0$ and $\Xi^-$ resonances into $\Lambda$ hyperons. 
Therefore, the measured $\pL$ correlation function includes residual correlations induced by the $p\Sigma^0$, $p\Xi^0$ and $p\Xi^-$ interactions. 
Such contributions, denoted as $C_{\pL_{\Sigma^0}}(k)$ and $C_{\pL_{\Xi}}(k)$, are obtained from chiral~\cite{HAIDENBAUER201324} and lattice calculations~\cite{SASAKI2020121737}, respectively. Other contributions are sub-dominant and considered as a flat correlation $C_\mathrm{flat} \approx 1$. 

The corrected correlation function is then given by the formula
\begin{eqnarray}\hspace*{-8mm}
	C_{\pL}(k)&=& B(k) \left[ \lambda_{\pL} C^\mathrm{th}_{\pL}(k) + \lambda_\mathrm{\pL_{\Sigma^0}} C_{\pL_{\Sigma^0}}(k) \right. \nonumber \\ & & \left.  \hspace*{2.5cm}
    + \lambda_{\pL_{\Xi}} C_{\pL_{\Xi}}(k) + \lambda_\mathrm{flat} \right],
	\label{eq:cplc}
\end{eqnarray}
where the $\lambda_i$ parameters represent the weights of the different contributions and are given in Table 1 of Ref.~\cite{pLambda}. The term $B(k)$ accounts for non-femtoscopic correlations, typically described by a polynomial function. In the region $k < 200$ MeV/$c$, these contributions are flat, hence, the function $B(k)$ has been considered as a normalization constant which is fixed by fitting the correlation function at low relative momenta. As the calculations do not include the effect of the $\pL$ to $N\Sigma$ coupling which is evident in the data at $k\approx 290$ MeV/c and do not include the contributions from $p-$ and $d-$waves, the fit is performed in the region $k < 120$ MeV/$c$, where such contributions are negligible. 
The comparison to the experimental correlation function is carried out by considering variations of the source radius and $\lambda_i$ parameters within the experimental uncertainties. The best agreement with the data, evaluated from the minimum value of the reduced $\chi^2$, is found for $r_0 = 1.19$~fm, $\lambda_{\pL} = 0.427$, $\lambda_{\pL_{\Sigma^0}} = 0.170$, $\lambda_{\pL_{\Xi}} = 0.221$ and $\lambda_\mathrm{flat} = 0.006$.

\begin{table}[]
    \centering
    \begin{tabular}{c|c}
      Model   &  $n_{tot}$ ($k<120$ MeV/c) \\
      \hline
        NLO13 (450) & 10.4\\
        NLO13 (500) & 9.1 \\
        NLO13 (550) & 8.0 \\
        NLO13 (600) & 8.7 \\
        NLO13 (650) & 8.6 \\
        NLO13 (700) & 8.3 \\
        NLO19 (500) & 7.7 \\
        NLO19 (550) & 6.9 \\
        NLO19 (600) & 6.5 \\
        NLO19 (650) & 6.5 \\
        N2LO (500) & 7.5 \\
        N2LO (550) & 7.8 \\
        N2LO (600) & 7.9 \\
        Bodmer-Usmani & 8.2 \\
        \hline
    \end{tabular}
    \caption{Number of standard deviations $n_{tot}$ between the theoretical and experimental correlation functions in the relative momentum region $k < 120$ MeV/c. The value of the cutoff $C$ for the NLO13, NLO19, N2LO Gaussian models is reported in brackets.}
    \label{tab:nSigma}
\end{table}

In Fig.~\ref{fig1}b the the genuine $\pL$ contribution $C_{\pL}^\mathrm{th}(k)$ (shown in Fig. 1a) has been scaled by the corresponding weight ($\lambda_{\pL}=0.427$) and combined with the other components by following Eq. \eqref{eq:cplc}. As seen from the figure, all the potential models remain close to each
other forming a narrow band. The Bodmer-Usmani potential is in the lower part of the band shown by the black dashed line. The lower panel in Fig.~\ref{fig1}b shows the level of agreement between the theoretical and the experimental correlation functions in each bin, evaluated in number of standard deviations $n_\sigma^i = (C_{\pL}^i- C_\mathrm{data}^i)/\sigma^i_\mathrm{data}$, where $\sigma^i_\mathrm{data}$ is the combined statistical and systematic uncertainties of the data in the $i-$th bin. The agreement in the region $k<120$ MeV/c is obtained from the total $\chi^2$ in the first 10 bins, which is converted into a total number of standard deviations ($n_{tot}$) from the p-value of the $\chi^2$ distribution. The values are reported in Table \ref{tab:nSigma}.

Among the considered models, the Gaussian representation of the NLO19 potential with cutoff at 600-650 MeV provide a better description of the experimental correlation function, as previously found from the analysis in Ref. \cite{pLambda}. 
The values in 
Table~\ref{tab:nSigma} are larger than those reported in \cite{pLambda} due to the different value adopted for the source radius, which has been recently corrected in Ref. \cite{ALICEsourceErr}. Although the correlation functions for the different models are very close to each other, they are not capable to describe the femtoscopy data with enough accuracy. Indeed, the deviation from the experimental correlation function is larger than 6.5$\sigma$. This motivates further analyses to improve the current understanding on the $\pL$ low energy interaction.

Variation of the potential parameters to see the impact in the correlation function has been done in the past~\cite{wang1999}. Very recently the femtoscopy data were incorporated in a simultaneous fit of the $\pL$ correlation function in different $m_T$ intervals, together with $p\Lambda$ scattering data, to constrain the scattering parameters~\cite{mihaylov2024}. 
The resulting parameters were at variance with respect to previous results. A set of scattering legths and effective ranges  describing both set of data is, for the singlet state $a_0=-2.10\,$fm and $r_e^0=3.21\,$fm,
and, for the triplet state $a_1=-1.54\,$fm and $r_e^1=3.16\,$fm. From these parameters a Gaussian representation can be constructed with strength and range in the singlet state of $V_0=-25.6\,$MeV and $r_0=1.485\,$fm and strength and range in the triplet state of $V_1=-27.55\,$MeV and $r_1=1.358\,$fm. Moreover, using this Gaussian potential the hypertriton binding energy results $B(^3_\Lambda{\rm H})=2.40\,$MeV, very close to the experimental value.
The correlation function constructed from such scattering parameters~\cite{mihaylov2024} is shown in Fig.~\ref{fig1}b bottom panel with the dashed (blue) line and indicated as Gauss Ref.~\cite{mihaylov2024}. The level of agreement with the data significantly improves, with $n_{tot} = 3.6$ in the region $k<120$ MeV/c. 
The idea behind these works is the use of the femtoscopy data in the study of the $\pL$ interaction. Following this idea, in the next sections we study the $\ppL$ correlation function to see if the available (or future) data could be used to study the $NN\Lambda$ interaction.

\section{Three-particle correlation function} \label{chap:threebodycase}
\label{sect3}

\subsection{Hyperspherical coordinates}
Three-body systems are usually described using the Jacobi coordinates, commonly denoted by $\bm{x}$ and $\bm{y}$. For two nucleons in positions (1,2) and the $\Lambda$ particle in position 3 we define them as
\begin{equation}
	\left\{ \begin{array}{l}
		\bm{x}=\bm{r}_2-\bm{r}_1 \\
		 \\
		\bm{y}=\sqrt{\frac{4}{(1+2m/M)}}\
		(\bm{r}_3-\frac{\bm{r}_1+\bm{r}_2}{2}) \; ,
                 \end{array}
		 \right. 
\end{equation}
where $\bm{r}_i$ is the position vector of particle $i$, $m$ is
the nucleon mass and $M$ is the mass of the $\Lambda$ hyperon.
The Jacobi coordinates are combined to construct the hyperspherical coordinates,
which contain one radial coordinate, the hyperradius $\rho=(x^2+y^2)^{1/2}$, and
five hyperangles (the four angles describing the directions of $\bm{x}$ and $\bm{y}$ plus
$\alpha=\arctan(x/y)$) that we collect into $\Omega_\rho$. 

With these definitions we have that
\begin{equation}
\frac{\rho^2}{2}=r_1^2+r_2^2+\frac{M}{m}r_3^2-\frac{M+2m}{m} R^2,
\label{relcoor}
\end{equation}
where $\bm{R}$ is the center of mass coordinate, and which in the center of mass frame can be rewritten as
\begin{equation}
	\rho^2=2 \left(r_1^2+r_2^2+\frac{M}{m} r_3^2\right)=
	\frac{2M}{M+2m}(r_{31}^2+r_{32}^2+\frac{m}{M} r_{12}^2)
\end{equation}
with $r_{ij}=|\bm{r}_i-\bm{r}_j|$ the relative distance between particles $i$ and $j$.

From the $\bm{x}$ and $\bm{y}$ coordinates the conjugate momenta, $\bm{k}$ and $\bm{q}$, can be defined:
\begin{equation}\label{eq:Jacobi}
	\left. \begin{array}{l}
		\bm{k}=\frac{1}{2}(\bm{p}_2-\bm{p}_1) \\
		 \\
		\bm{q}=\sqrt{\frac{m}{M}}\sqrt{\frac{m}{2m+M}}\
		(\bm{p}_3-\frac{M}{m}\frac{\bm{p}_1+\bm{p}_2}{2}),
                 \end{array}
		 \right. 
\end{equation}
where $\bm{p}_i$ is the momentum of particle $i$. We can also construct the hypermomentum
$Q=(k^2+q^2)^{1/2}$ and the five hyperangles $\Omega_Q$ equivalent to $\Omega_\rho$,
but in momentum space. 

In the center of mass system the following relations hold:
\begin{eqnarray}
\lefteqn{
	Q^2=\frac{1}{2} \left(p_1^2+p_2^2+\frac{m}{M} p_3^2\right)=}
 \nonumber \\ &&
	\frac{(M+m)^2}{2M(M+2m)}\left[p_{31}^2+p_{32}^2+\frac{4Mm}{(M+m)^2} p_{12}^2\right],
\end{eqnarray}
with $p_{ij}$ the modulus of the relative momentum defined as
$(m_j\bm{p}_i-m_i\bm{p}_j)/(m_i+m_j)$. In the particular case considered here in
which the two nucleons are in positions 1 and 2, and the $\Lambda$ in position 3,
they take the form: $\bm{p}_{12}=(\bm{p}_1-\bm{p}_2)/2$,
$\bm{p}_{23}=(M\bm{p}_2-m\bm{p}_3)/(m+M)$ and 
$\bm{p}_{31}=(m\bm{p}_3-M\bm{p}_1)/(m+M)$.
The total energy of the system in the center of mass frame is
\begin{equation}
	E=\frac{p_1^2}{2m}+\frac{p_2^2}{2m}+\frac{p_3^2}{2M}=\frac{Q^2}{m} \ .
\end{equation}

These coordinates are the ones employed in the hyperspherical harmonic (HH)
formalism~\cite{HH2008,HH2020}.
In the experimental measurements, three-particle correlation functions are defined in terms of a Lorentz invariant quantity $Q_3$. In the non-relativistic limit it is defined as~\cite{projector}
\begin{equation}
    Q_3^2=4\ (p^2_{12}+p_{23}^2+p_{31}^2) \ ,
    \label{Q3a}
\end{equation}
which is, in general, not equal to the hypermomentum $Q$ used in the calculations. 

The kinematic variable $Q_3$ is related to the Jacobi vectors defined in Eq.~\eqref{eq:Jacobi} as
\begin{equation}
    \begin{aligned}[b]
        &
        Q_3^2 = \left[ \frac{8M^2}{(m + M)^2} + 4 \right] \bm{k}^2 +
        \left[ \frac{8M(M+2m)}{(m + M)^2} \right] \bm{q}^2 \ ,
    \end{aligned}
    \label{Q3b}
\end{equation}
which for $m=M$ becomes $Q_3^2 = 6 Q^2$. When $m$ and $M$ refer, respectively, to the proton and $\Lambda$ masses, we have $Q^2=6.36 \bm{k}^2+6.33 \bm{q}^2$, which can
be approximated to $Q_3^2 \approx 6.35 Q^2$. This last relation will be used in the following.

\subsection{The source function}

Similarly to Eq.~(\ref{cor1e2}), the correlation function for three particles is given by
\begin{equation}
 C_{123}(Q) =
 \int d\bm{r}_1 d\bm{r}_2 d\bm{r}_3\; S_1(\bm{r}_1) S_2(\bm{r}_2) S_3(\bm{r}_3) |\Psi_s|^2\ ,
	\label{eq:c123}
\end{equation}
where $Q$ is the hypermomentum and $\Psi_s$ the three-particle scattering wave
function. 

The source function $S_i(\bm{r}_i)$ for particle $i$ is again approximated as the Gaussian probability distribution given in Eq.(\ref{spf12}), where we now denote the widths of the proton and $\Lambda$ distributions as $R_m$ and $R_M$, respectively. In this way the product of the three source functions takes the form
\begin{eqnarray}
\lefteqn{
 S_1(\bm{r}_1) S_1(\bm{r}_2) S_1(\bm{r}_3) =} \nonumber \\ &&
	= \frac{1}{(2\pi R_m^2)^{3}} \frac{1}{(2\pi R_M^2)^{\frac32}}
 e^{-(r_1^2+r_2^2)/2R_m^2-r_3^2/2R_M^2} 
 \nonumber \\ &&
	=  \frac{e^{-(\frac{\rho^2}{2}-(\frac{R_m^2}{R_M^2}-\frac{M}{m})r_3^2+\frac{M+2m}{m}R^2)/2R_m^2}}{(2\pi R_m^2)^{3} (2\pi R_M^2)^{\frac32}  } \ ,
 \label{totsour}
\end{eqnarray}
where Eq.~(\ref{relcoor}) has been used.

Since the three particles are not identical, we obtained a non hypercentral total source function. Indeed, the hypercentrality of the source would require a specific relation between the particle masses and source radii, i.e., $R_m^2/R_M^2=M/m$. 

The single particle radii $R_m$ and $R_M$ can be extracted from the effective radii of $NN$ and $N\Lambda$ pairs within the $NN\Lambda$ triplets, using the relation in Eq.~\eqref{eq:r0}. In the $\ppL$ data sample measured by ALICE~\cite{femtoppp}, the transverse mass of the $\pp$ and $\pL$ pairs in the triplets are $\langle m_T\rangle^{pp} = 1.30$ GeV/$c$ and $\langle m_T\rangle^{\pL} = 1.43$ GeV/$c$, respectively. 
The corresponding effective Gaussian radii for $\pp$ and $\pL$ pairs, obtained from the source model in Refs.~\cite{ALICEsource,ALICEsourceErr}, are
$r_{0}^{pp} = (1.31 \pm 0.07) \text{ fm}$ and $r_{0}^{\pL} = (1.30 \pm 0.07) \text{ fm}$.
The uncertainties are propagated from the measurement of the reference $\pp$ core radii, by including 10\% variations on the resonance yields in the adopted source model. 
Combining the effective radii in Eq.~\eqref{eq:r0}, we get
\begin{equation}\label{eq:SPSource}
    \begin{aligned}[b]
        &
        R_{m}= (1.31 \pm 0.07) \text{ fm} \\
        &
        R_{M} = (1.30 \pm 0.16) \text{ fm} \ .
    \end{aligned}
\end{equation}

Due to the fact that $m$ and $M$ are not very different and the source radii
are similar, $R_m\approx R_M$, in the following we
assume that the relation $R_m^2/R_M^2=M/m$ is sufficiently fulfilled. In this way the total source function (\ref{totsour}) takes the form:
\begin{equation}
 S_1(\bm{r}_1) S_1(\bm{r}_2) S_1(\bm{r}_3)
	= \left(\frac{M}{m}\right)^{\frac{3}{2}}
 \frac{e^{-(\frac{\rho^2}{2}+\frac{M+2m}{m}R^2)/2R_m^2}}
	{(2\pi R_m^2)^{\frac92}} \ .
\end{equation}

As done in Eq.~(\ref{tpf12}), integrating out the center of mass coordinate, the source function has the hypercentral form
\begin{equation}
S_{123}(\rho)=\frac{1}{\pi^3\rho_0^6}e^{-(\rho/\rho_0)^2}\ , \label{sour3b}
\end{equation}
with $\rho_0=2R_m$ and the normalization condition
\begin{equation}
\int S_{123}(\rho) \rho^5 d\rho \,d\Omega_\rho=1,
\end{equation}
leading to the following final expression for the three-body correlation 
function:
\begin{equation}
    C_{123}(Q)=\frac{1}{\pi^3 \rho_0^6}
    \int e^{-(\rho/\rho_0)^2} |\Psi_s|^2 \rho^5 d\rho d\Omega_\rho.
    \label{finc}
\end{equation}

Although the sensitivity of the correlation function on the source radius $\rho_0$ will be investigated, taking into account the $R_m$ value given in Eq.~(\ref{eq:SPSource}) and the fact
that $\rho_0=2R_m$, most of the results shown in the following sections, and in particular the comparison to the ALICE data, will be done considering $\rho_0 = 2.6$~fm. 

\subsection{The three-body continuum wave function}

Within the HH method the three-body scattering wave function can be written as
\cite{gar14}
\begin{eqnarray}
\lefteqn{
    \Psi_s=\frac{1}{\sqrt{N_S}}\frac{(2\pi)^3}{(Q\rho)^{5/2}} }
    \label{3bdcon} \\ && \times
    \sum_{JJ_z} \sum_{K\gamma}\Psi_{K\gamma}^{J J_z}
    \sum_{M_LM_S} (LM_L SM_S|JJ_z) {\cal Y}_{KL M_L}^{\ell_x\ell_y}(\Omega_Q)^*, 
    \nonumber
\end{eqnarray}
where $N_S$ is the number of spin states and $\gamma$ groups the quantum numbers $\{\ell_x,\ell_y,L,s_x,S\}$, 
$\ell_x$ and $\ell_y$ are the relative orbital angular momenta associated to 
the Jacobi coordinates $\bm{x}$ and $\bm{y}$ coupled to total orbital angular 
momentum $L$ (with projection $M_L$). The spin $s_x$ denotes the spin of the two nucleons
connected by the $\bm{x}$ coordinate, which couples to the spin of the third particle to give
the total three-body spin $S$ (with projection $M_S$). The angular momenta $L$
and $S$ are coupled to the total angular momentum of the system $J,J_z$.
Finally, $K=2\nu+\ell_x+\ell_y$ (with $\nu=0,1,2,\cdots$)  is the grand-angular momentum quantum number, and
${\cal Y}_{KLM_L}^{\ell_x \ell_y}$ are the usual hyperspherical harmonic functions.

The coordinate wave functions, $\Psi_{K\gamma}^{J J_z}$, take the general form
\begin{equation}
    \Psi_{K\gamma}^{J J_z}=\sum_{K'\gamma'} \Psi^{K'\gamma'}_{K\gamma}(Q,\rho) \Upsilon_{JJ_z}^{K'\gamma'}(\Omega_\rho)\ ,
    \label{3bdcoo}
\end{equation}
with
\begin{equation}
    \Upsilon_{JJ_z}^{K\gamma}(\Omega_\rho)=\sum_{M_L M_S} (L M_L S M_S |J J_z) 
    {\cal Y}_{KLM_L}^{\ell_x\ell_y}(\Omega_\rho) \chi_{SM_S}^{s_x}.
    \label{upsilon}
\end{equation}

The radial wave function, $\Psi^{K'\gamma'}_{K\gamma}$, corresponds to 
a process with incoming and outgoing channels characterized by the set of quantum
numbers $\{K,\gamma\}$ and $\{K',\gamma'\}$, respectively.

Following Eq.~(\ref{norm}), we define again the norm of the scattering wave function as the average over
the angular coordinates of the square of the wave function, i.e.:
\begin{equation}
    |\Psi_s|^2_\Omega = \frac{1}{\pi^6} \int d\Omega_\rho \int d\Omega_Q |\Psi_s|^2.
    \label{norm3b}
\end{equation}

In the case of non-interacting particles, we have that \cite{gar14}
\begin{equation}
    \Psi^{K'\gamma'}_{K\gamma}(Q,\rho)=i^K \sqrt{Q\rho}J_{K+2}(Q\rho)\delta_{KK'}\delta{\gamma \gamma'},
    \label{prad}
\end{equation}
where $J_{K+2}(Q\rho)$ is the Bessel function of order $K+2$,
and the continuum wave function of Eq.(\ref{3bdcon}) becomes
\begin{equation}
\Psi_s^0=\frac{1}{\sqrt{N_S}}e^{i\bm{Q}\cdot \bm{\rho}} \sum_{S M_S s_x} \chi_{SM_S}^{s_x}.
\label{free3b}
\end{equation}

Using the partial wave expansion
\begin{eqnarray}
\lefteqn{
    e^{i\bm{Q}\cdot\bm{\rho}}=e^{i(\bm{k}\cdot \bm{x}+\bm{q} \cdot \bm{y} )}=} \label{3bdpw} \\ &&  \hspace*{-4mm}
    \frac{(2\pi)^3}{(Q\rho)^2}
\sum_{K\ell_x\ell_y L M_L} i^K J_{K+2}(Q\rho) 
{\cal Y}_{KL M_L}^{\ell_x\ell_y}(\Omega_\rho) {\cal Y}_{KL M_L}^{\ell_x\ell_y}(\Omega_Q)^*, \nonumber
\end{eqnarray}
we can verify that the norm of the three-body plane wave, Eq.~(\ref{norm3b}), is given by 
\begin{equation}
\frac{16}{3(Q\rho)^4} \sum_{K} J_{K+2}^2(Q\rho)(K+3)(K+2)^2(K+1)=1,  \,
\end{equation}
where we have used the following property of the HH functions:
\begin{equation}
\sum_{\ell_x\ell_y L M_L} {\cal Y}_{KLM_L}^{\ell_x\ell_y*} {\cal Y}_{KLM_L}^{\ell_x\ell_y}=\frac{1}{12\pi^3}(K+3)(K+2)^2(K+1).
\label{eq:HHproperty}
\end{equation}

Insertion of Eq.~(\ref{free3b}) into Eq.~(\ref{norm3b}) trivially leads to the same result as in two-body case, i.e., that the norm of the free scattering function is $|\Psi^0_s|^2_\Omega=1$, where now $N_S=8$ (4 spin states from $S=\frac{3}{2}$, 2 from $S=\frac{1}{2}$ with $s_x=0$, and 2 from $S=\frac{1}{2}$ with $s_x=1$).

\subsection{Antisymmetrization of the $NN\Lambda$ wave function}

The three-body wave function, as given in Eq.(\ref{3bdcoo}), does not have a well-defined symmetry under particle exchange. In the $NN\Lambda$ system the wave function has to be antisymmetric in the coordinates of the two nucleons. 
This means that, if the spin function with total spin $S$ is fully symmetric (antisymmetric) under exchange of the two identical nucleons, only antisymmetric (symmetric) HH functions are allowed in the expansion in Eq.(\ref{3bdcoo}). In the same way, if the spin function with total spin $S$ mixes symmetric and antisymmetric states, the HH states entering in the expansion  (\ref{3bdcoo}) have to correspondingly mix antisymmetric and symmetric states
such that the total antisymmetry is preserved.

To be precise, this last situation is what happens when the total spin 
of the $NN\Lambda$ system is $S=\frac{1}{2}$. The spin states are in this case given by
\begin{equation}
\chi^{\lambda}_{S=\frac{1}{2}S_z}=\sum_{\sigma_x\sigma_y} (\lambda \sigma_x \, \frac{1}{2} \sigma_y |S=\frac{1}{2}S_z) \chi_{\lambda \sigma_x} 
\chi_{\frac{1}{2}\sigma_y},
\label{sstate}
\end{equation}
where $\chi_{\lambda\sigma_x}$ and  $\chi_{\frac{1}{2}\sigma_y}$ are, respectively, the spin functions of the two-body system formed by the nucleons 1 and 2, and the one of the $\Lambda$-particle. The quantum number $\lambda=0,1$ corresponds to the antisymmetric or symmetric spin state with respect to the exchange of particles $1,2$, respectively. 
Since the total wave function has to be antisymmetric in particles (1,2), we can rewrite Eq.(\ref{upsilon}) for $S=\frac{1}{2}$ as
\begin{equation}
\Upsilon_{JJ_z}^{K\gamma }(S=\frac{1}{2})= \sum_{M_L M_S} (L M_L \frac{1}{2} M_S |J J_z) 
    {{\cal Y}_{KLM_L}^{\ell_x\ell_y,\bar{\lambda}}(\Omega_\rho) \chi^\lambda_{\frac{1}{2}M_S}} \ ,
    \label{eq:mixs}
\end{equation}
where we have introduced the HH functions ${\cal Y}_{KLM_L}^{\ell_x\ell_y,\bar{\lambda}}$ having well defined values of angular momentum $LM_L$ and symmetry $\bar\lambda$, either symmetric ($\bar{\lambda}\equiv s$) or
antisymmetric ($\bar{\lambda}\equiv a$), under exchange of the two identical nucleons. Therefore, since the total wave function has to be antisymmetric in particles (1,2), the $S=\frac{1}{2}$ terms in the expansion (\ref{3bdcoo}) mix symmetric and antisymmetric HH states, in such a way that $\bar\lambda\equiv a$ when $\lambda=1$ and $\bar\lambda\equiv s$ when $\lambda=0$.

In the case of $S=\frac{3}{2}$ the spin part is always symmetric under exchange of nucleons 1 and 2, and we have that 
\begin{equation}
    \Upsilon_{JJ_z}^{K\gamma}(S=\frac{3}{2})=\sum_{M_L M_S} (L M_L \frac{3}{2} M_S |J J_z)
    {\cal Y}_{KLM_L}^{\ell_x\ell_y,\bar\lambda}(\Omega_\rho) \chi^1_{\frac{3}{2}M_S} \ ,
\end{equation}
with $\bar\lambda\equiv a$, and only antisymmetric HH functions enter in 
Eq.(\ref{3bdcoo}).
The consequence of imposing the required antisymmetry under exchange of the
two identical nucleons is that in both cases, $S=\frac{1}{2}$ and $\frac{3}{2}$, 
the relative orbital angular momentum between the two nucleons, $\ell_x$, and the two-nucleon spin, $s_x$, are not independent. The even (odd) character of $\ell_x$ imposes $s_x=0$ ($s_x=1$), and therefore the index $\gamma$ in the summation in Eq.(\ref{3bdcoo})
does not include anymore the value of $s_x$, which is fixed by the symmetry requirements.

\subsection{The $NN\Lambda$ Three-Body Force}\label{sec:NNLforce}

At this stage it is important to consider the impact on the correlation function when a three-body force is
included in the description of the $NN\Lambda$ system. From a theoretical
point of view, the necessity of inclusion of a three-body force 
is well justified within the EFT framework. Also, the description of the hypernuclei
using only two-body forces is not completely satisfactory, since the binding energies of many of the light hypernuclei, calculated using two-body forces, are at variance with
the experimental results. This is similar to what is observed in nuclear
physics, where nuclear binding energies cannot be described with only two-body nuclear
forces. In nuclear physics the three-nucleon force has to provide attraction in the
light nuclear sector whereas, as the nuclear density increases, it has to
prevent the nuclear collapse providing a short-range repulsion. In the case of
hypernuclei, a repulsive three-body force seems to be necessary.
To be noticed that a $NN\Lambda$ three-body force could have a prominent role in
the description of compact systems. In fact not all theory models are in
agreement with the observation of neutron stars having a mass equivalent to two
solar masses or even higher. 

Within the chiral EFT, three-body forces appear as subleading terms in the
expansion. Conversely, using an EFT based on contact interactions, the three-body
force appears at leading order. An example of this kind of EFT is used to
describe systems with large values of the scattering length~\cite{report_hammer}. 
The system is close to the unitary limit, a limit in which the two-body system
has a bound state at zero energy. The treatment of the three-body
system within this theory shows that this system is unstable (Thomas collapse~\cite{thomas}) requiring the
introduction of a three-body force~\cite{bedaque}.  
When this theory is applied in nuclear physics it is called pionless
EFT~\cite{vankolck,rocco}.  
In this context the Gaussian interactions introduced in Section~\ref{sect2} represent a leading order description
in a contact EFT and the Hamiltonian of the system, considered at that order, has to be
completed by introducing a three-body force. 

Here we will use a three-body force of the form
\begin{equation}
	 W(r_{13},r_{23})= W_3 \, e^{-(r_{13}^2+r_{23}^2)/\rho_3^2} \ ,
\end{equation}
where $r_{13}$ and $r_{23}$ are the distances between the $\Lambda$-particle and
each of the two nucleons, and where $W_3$ and $\rho_3$ are parameters fixed to describe 
as good as possible the
binding energies of the lightest hypernuclei having a single $\Lambda$ particle.

In Table~\ref{Tab2} we report the binding energy of the hypertriton, $^3_\Lambda$H,
using the Gaussian two-body forces (see the row labeled by $B(^3_\Lambda{\rm H})$). A consistent
overbinding can be observed considering that the experimental value of the
$\Lambda$ separation energy is $B_\Lambda=0.164\pm0.043\,$keV 
(see https://hypernuclei.kph.uni-mainz.de). Accordingly the binding energy of
the hypertriton results $B(^3_\Lambda{\rm H})\approx 2.39\,$MeV. The last two
rows of Table~\ref{Tab2} give the strength and range of the three-body force, which, 
associated with the corresponding two-body force,
reproduces this value. To be noticed that the value of the range has been selected to give also a good description of the binding energies of the two four-body hypernuclei, $^4_\Lambda$H and $^4_\Lambda$He. 
As already discussed in Section \ref{sec:Results2B}, the hypertriton binding energy obtained with the Gaussian model from Ref.~\cite{mihaylov2024}, which provides the best description of both scattering and femtoscopy data, is close to the experimental value and the overbinding of such hypernucleus is significantly reduced. Further investigations with this model will be addressed in a future work.
To calculate the binding energy of the three- and four-body system with one $\Lambda$ we use two methods, the HH basis expansion without permutation symmetry~\cite{gatto2009,nannini2018} and the stochastic variational method as described in Ref.~\cite{gattobigio2019}.

\section{The $\nnL$ system}
\label{sect4}

Considering the antisymmetrization of particles (1,2), we obtain that
the norm of the free scattering state, Eq.~(\ref{norm3b}), becomes
\begin{equation}
|\Psi^0_s|^2_\Omega = \frac{2}{N_S}\frac{2^6}{(Q\rho)^4} \sum_{K} J_{K+2}^2(Q\rho) N_{ST}(K)\ ,
\label{free3n}
\end{equation}
where the factor of 2 ($=2!$) enters due to the fact that we are dealing
with two identical particles, the factor $1/N_S$, introduced in Eq.(\ref{3bdcon}), guarantees that $|\Psi^0_s|^2_\Omega\rightarrow 1$ as $Q\rho\rightarrow \infty$, and $N_{ST}(K)$ is the number of states consistent with the value of the grand-angular quantum number $K$.

To calculate $N_{ST}(K)$ we have to consider that for each value
of the grand-angular quantum number, $K$, the HH functions can be either fully symmetric, fully antisymmetric, or mixing both symmetries (mixed HH functions) with respect to the exchange of the three particles. Concerning the spin states, there is no fully antisymmetric state for three spin-$1/2$ particles, and the spin vector can only be  either of mixed symmetry if $S=\frac{1}{2}$ or symmetric if $S=\frac{3}{2}$. 

In any case, the mixed symmetry spin states having $S=\frac{1}{2}$ can have $\lambda=0$ in Eq.(\ref{sstate}), being antisymmetric under exchange of the two nucleons. Therefore, in order to get the correct antisymmetry under exchange of particles 1 and 2 in the full wave function, these spin states have to combine with either fully symmetric HH functions or mixed HH functions but symmetric in particles 1 and 2 ($\bar{\lambda}\equiv s$ in Eq.(\ref{eq:mixs})). In the same way, the mixed symmetry spin states with $S=\frac{1}{2}$ and $\lambda=1$ are symmetric in particles 1 and 2, and they must then combine with fully antisymmetric HH functions or mixed HH functions but antisymmetric in particles 1 and 2 ($\bar{\lambda}\equiv a$ in Eq.(\ref{eq:mixs})). Finally, the fully symmetric spin state with $S=\frac{3}{2}$ combines with either the fully antisymmetric HH states or the mixed HH functions with $\bar{\lambda}\equiv a$.

As a result of the discussion above we therefore get that the norm is
\begin{eqnarray}
\lefteqn{
|\Psi^0_s|^2_\Omega = } \label{free3na} \\ && \hspace*{-3mm}
\frac{1}{4}\frac{2^6}{(Q\rho)^4} \sum_{K} J_{K+2}^2(Q\rho)
	( 2N^T_{ST}(K)+2N^m_{ST}(K)+4N^a_{ST}(K)) \ ,
\nonumber
\end{eqnarray}
with $N_{ST}^T(K)$ the total number of HH states, $N_{ST}^m(K)$ the number of mixed HH functions and 
$N_{ST}^a(K)$ the number of antisymmetric HH functions
for each value of $K$. The factors in front of $N_{ST}^T(K)$, $N_{ST}^m(K)$ and
$N_{ST}^a(K)$ correspond to the spin degeneracy. The algorithm used 
to determine the number of HH functions having different symmetries is given in
Ref.~\cite{pppth} (see also Ref.~\cite{gatto2009}). As a comment on the formula, the number of symmetric and antisymmetric states, $N_{ST}^s(K)$, $N_{ST}^a(K)$, is very similar as $K$ increases, and therefore the number of states asymptotically results in
$2N^T_{ST}(K)+2N^m_{ST}(K)+4N^a_{ST}(K)\rightarrow 4N^T_{ST}(K)$,
and the norm tends to unity.

Turning back to the three-body correlation function given in Eq.~(\ref{finc}),
due to the spherical symmetry of the source function, we can replace $|\Psi_s|^2$, 
given in Eq.~(\ref{3bdcon}), by $|\Psi_s|^2_\Omega$, as defined in Eq.~(\ref{norm3b}).
In the particular case of considering the $\nnL$ system as made of three non-interacting particles, and making use of Eq.~(\ref{free3n}) with $N_S=8$, we then get 
\begin{equation}
C_{123}(Q)= \frac{1}{4}\frac{2^6}{Q^4\rho_0^6}
\int \rho\, d\rho\, e^{-\frac{\rho^2}{\rho_0^2}}\, \sum_K J^2_{K+2}(Q\rho) N_{ST}(K).
\label{cnnn}
\end{equation}

\subsection{Introducing the interaction}

The analytical form of the three-body continuum wave function given in Eqs.~(\ref{3bdcon}) and (\ref{3bdcoo}) 
is completely general. In the case of non-interacting particles the matrix formed by the radial wave functions, 
$\Psi_{K\gamma}^{K'\gamma'}$, is diagonal. When a short-range interaction is
present, this matrix could be non-diagonal with different sets of quantum numbers $\{K,\gamma\}$ and
$\{K',\gamma'\}$ characterizing the incoming and outgoing channels. 
Within the HH formalism the basis set used in the wave function expansion
could be quite large leading to a very large system of coupled radial equations 
to be solved. As we have discussed in Ref.~\cite{pppth}, in the case of the $\nnn$ and $\ppp$
correlation function, it is convenient to employ the hyperspherical adiabatic
(HA) basis, denoted by $\Phi_n^{JJ_z}(\rho,\Omega_\rho)$,
where $n$ labels the different basis elements depending not only on the hyperangles, $\Omega_\rho$,
but also on the hyperradius, $\rho$. The HH and HA basis elements are related by
the following transformation
\begin{equation}
    \Phi_n^{JJ_z}(\rho,\Omega_\rho) = \sum_{K\gamma} \langle \Upsilon_{JJ_z}^{K\gamma}(\Omega_\rho) | 
    \Phi_n^{JJ_z}(\rho,\Omega_\rho) \rangle_{\Omega_\rho}    \Upsilon_{JJ_z}^{K\gamma}(\Omega_\rho),
    \label{bastr}
\end{equation}
where the functions $\Upsilon_{JJ_z}^{K\gamma}(\Omega_\rho)$ are given in Eq.(\ref{upsilon}), and 
$\langle  |  \rangle_\Omega$ represents integration over the angular
coordinates. In practical applications the summation 
over the HH quantum numbers $K$ and $\gamma\equiv\{\ell_x\ell_y L S \}$ has to be truncated at some maximum values
$K_\mathrm{max}$ and $\gamma_\mathrm{max}$.

Using the HA basis, the three-body continuum wave function can be written as
(see the Appendix D of Ref.~\cite{gar14})
\begin{equation}
    \Psi_s = \frac{(2\pi)^3}{(Q\rho)^{5/2}} \sum_{JJ_z} \sum_n u_n^J 
    \sum_{s_x S M_S} \langle \chi_{SM_S}^{s_x} | \Phi_n^{JJ_z} (Q,\Omega_Q) \rangle^*,
\label{3bdad}
\end{equation}
where 
\begin{equation}
    u_n^J= \sum_{n'} u_n^{n'}(Q,\rho) \Phi_{n'}^{JJ_z}(\rho,\Omega_\rho),
    \label{3bdad2}
\end{equation}
and the incoming and outgoing channels are now $n$ and $n'$, respectively.
The HA basis is useful to reduce the number of coupled equations in the
determination of the hyperradial functions $u_n^{n'}(Q,\rho)$. In fact
most of the dynamics of the system is captured by the lowest terms in the adiabatic 
expansion. 

An important point to be mentioned is that each term of the HA basis tends at large distances to a single HH basis element term:
 \begin{equation}
	\Phi^{J J_z}_n(\rho,\Omega) 
 \stackrel{\rho\rightarrow \infty}{\longrightarrow} \Upsilon_{JJ_z}^{K\gamma}(\Omega)\ ,
 \label{hhad}
\end{equation}
allowing to associate a specific grand-angular momentum value, $K$, to each adiabatic channel $n$. Therefore, in the following, when referring to an adiabatic channel with a certain value of $K$ it is in the asymptotic sense indicated in Eq.(\ref{hhad}), since in the inner region many values of $K$ contribute (in the actual calculations all the $K\leq K_\mathrm{max}$ terms, see Eq.(\ref{bastr})).

The asymptotic behavior of the hyperradial functions, $u_n^{n'}(Q,\rho)$, have the form
\begin{eqnarray}
\lefteqn{
	u_n^{n'}(Q,\rho\rightarrow\infty) 
\equiv u_K^{K'}(Q,\rho\rightarrow\infty) 
 \rightarrow }
 \label{eq:asymu} \\ &&
		i^{K'}\sqrt{Q\rho} \left[ \delta_{KK'}J_{K'+2}(Q\rho)+ T_{KK'}^{\gamma\gamma'} {\cal O}_{K'+2}(Q\rho) \right],
\nonumber
\end{eqnarray}
where $K$ and $K'$ are the grand-angular momentum values associated to the incoming and
outgoing channels $n$ and $n'$, respectively, ${\cal O}_{K+2}(Q\rho)=Y_{K+2}(Q\rho)+i J_{K+2}(Q\rho)$ is the 
the outgoing asymptotic wave function,
$T_{KK'}^{\gamma\gamma'}$ is a $T$-matrix element, and $J_{K+2}(Q\rho),Y_{K+2}(Q\rho)$ are the regular and 
irregular Bessel functions. 

The norm of the scattering wave function for the three-body
state with total angular momentum $J$, tends asymptotically to
\begin{equation}
	|\Psi^{J}|^2_\Omega(\rho \rightarrow \infty)= 
 (2J+1) \frac{1}{4}\frac{2^6}{(Q\rho)^5}\sum_{KK'}
	|u_K^{K'}(Q,\rho\rightarrow\infty)|^2 ,
\end{equation}
where the $(2J+1)$ factor appears after
summation over all the possible $J_z$ projection quantum numbers.
After summation over all the possible $J$ states, the norm becomes
\begin{small}
\begin{eqnarray}
\lefteqn{
	\sum_{J} |\Psi^{J}|^2_\Omega= \frac{1}{4}\frac{2^6}{(Q\rho)^4} \times}
 \label{eq:normf} \\ && \hspace*{-4mm}
	\left( \sum_J (2J+1) \sum_{KK'}^{K_0}
	\left|\frac{u_ K^{K'}(Q,\rho)}{\sqrt{Q\rho}}\right|^2
	+ \sum_{K>K_0} J_{K+2}^2(Q\rho) N_{ST}(K) \right) \ ,
\nonumber
\end{eqnarray}
\end{small}
where $K_0$ is the quantum number indicating the maximum value of $K$ at which the interaction distorts
the free scattering state. For $K>K_0$ the wave function is taken as the free solution, 
i.e., for $K>K_0$,  $u_K^{K'}(Q,\rho)$ is replaced by
$i^{K'}\sqrt{Q\rho} \delta_{KK'} J_{K'+2}(Q\rho)$. In the above summation the
first sum includes all values of $J$ compatible with the value of $K\le K_0$ at which
the interaction has been considered. 

In the case of the $\nnL$ system the
sum in Eq.(\ref{eq:normf}) starts at $K=0$ which is compatible with the angular momentum and parity
values $J^\pi=1/2^+$. The next term corresponds to $K=1$ and it is compatible with
$J^\pi=1/2^-,3/2^-,5/2^-$. The advantage of using the HA basis is that 
the use of the lowest adiabatic channel to describe the scattering states is already a very good
approximation. In this case, Eq.~(\ref{3bdad2}) reduces to 
\begin{equation}
	u_{n_0}^J=  u_{n_0}(Q,\rho) \Phi_{n_0}^{JJ_z}(\rho,\Omega_\rho),
    \label{3bdad3}
\end{equation}
with $n_0$ the lowest adiabatic channel compatible with $J$. To get inside to this fact, the adiabatic channel $\Phi_{n_0}^{JJ_z}$ can be expanded in terms of HH functions as given in Eq.~{(\ref{bastr})}, which, as mentioned, has to be truncated
at some maximum values of the quantum numbers $K_\mathrm{max}$ and $\gamma_\mathrm{max}$, large enough to describe the dynamics of the system.

Using as well the lowest adiabatic channel to describe
the $J^\pi=1/2^+,1/2^-,3/2^-,5/2^-$ (which correspond to $K=1$) and free states for $K>1$, the $\nnL$ correlation function 
results
\begin{eqnarray}
\lefteqn{
C_{\nnL}(Q)= \frac{1}{4}\frac{2^6}{Q^4\rho_0^6} 
	\int \rho\, d\rho\, e^{-\frac{\rho^2}{\rho_0^2}} \times }
 \label{cnnL} \\ &&
 \left(\sum_J
	(2J+1)\left|\frac{u^J_{n_0}}{\sqrt{Q\rho}}\right|^2
	+  \sum_{K>1} J^2_{K+2}(Q\rho) N_{ST}(K)
	\right),
\nonumber
\end{eqnarray}
where the sum over $J$ includes the values $J^\pi=1/2^+,1/2^-,3/2^-,5/2^-$ which
exhausts the possibilities of $K=0,1$ as the possible asymptotic HH states.
Starting with $K=2$ the scattering process is considering as free. 
\subsection{The correlation function}

Let us now analyze the $\nnL$ correlation function using the Gaussian
representations of the $N\Lambda$ interaction given in Table~\ref{Tab2} (in this first analysis we assume
that the $\pL$ and $\nL$ interactions are the same). For the $nn$
interaction we use the $s$-wave Gaussian potential
\begin{equation}
	V_{\nn}= V_0 e^{-r^2/r_0^2}
\end{equation}
with $V_0=-30.45$ MeV and $r_0=1.815$ fm, fixed to reproduce the singlet $\nn$ scattering length,
$a_{\nn}=-18.9\pm 0.4$ fm, and effective range, $r_e=2.75\pm 0.11$ fm (see Ref.~\cite{machleidt2011}). The computed correlation functions will be shown as a function of $Q_3$, which can be related the three-body momentum $Q$ through Eqs~(\ref{Q3a}) and (\ref{Q3b}).

\begin{figure}[t!]
	\includegraphics[width=\columnwidth]{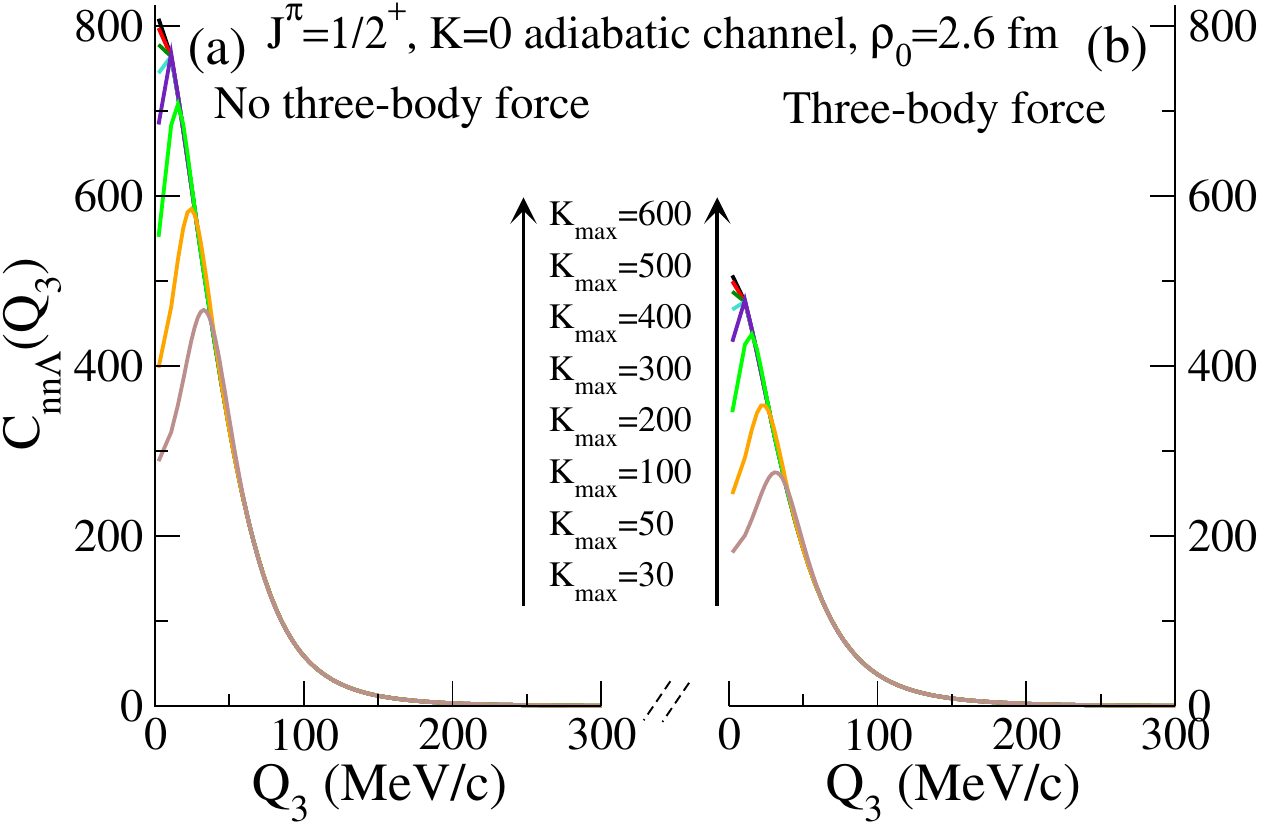}
	\caption{The $\nnL$ correlation function for the $1/2^+$ state including the lowest adiabatic potential ($K=0$) only. The source function, Eq.~(\ref{sour3b}), has been taken with $\rho_0=2.6$ fm. The $\nL$ interaction has been taken to be the Gaussian representation of the NLO19 potential with $C=600$ MeV as given 
 in Table~\ref{Tab2}. Panels (a) an (b) correspond to calculations without and with three-body force, respectively. The different curves are the calculations with
different maximum value of the grand-angular momentum $K$ used in the description
of the adiabatic potential, Eq.(\ref{bastr}). As indicated by the arrow in the figure, the higher $K_\mathrm{max}$ the higher the maximum of the correlation function. Convergence is reached for $K_\mathrm{max}\approx 500 $. }
	\label{fig:nnLKmax}
\end{figure}

We first study the $J^\pi=1/2^+$ state. Only the lowest adiabatic term in the expansion 
of Eq.(\ref{3bdad}) is considered, and the corresponding HA basis term, $\Phi_0^{J J_z}$, goes asymptotically to the $K=0$ HH state times the $S=1/2$ spin function. 
In Fig.~\ref{fig:nnLKmax} we investigate the convergence of the correlation function in terms of the number of HH functions needed to describe $\Phi_0^{JJ_z}$, that is, as a function of the $K_\mathrm{max}$ value used in the expansion of Eq.(\ref{bastr}). The $\nnL$ correlation function has a very high peak at
low energies and, to correctly describe it, a pretty large number of HH functions is needed. 
The convergence is particularly difficult close to the origin, where  values of $K_\mathrm{max}>500$ are needed. The study has been done
using the Gaussian representation of the NLO19 potential with cutoff $C=600\,$
MeV and a source radius of $\rho_0=2.6\,$fm. Fig.~\ref{fig:nnLKmax}a shows the results
without the use of any three-body force, and in Fig.~\ref{fig:nnLKmax}b the three-body force given in the last two rows of Table~\ref{Tab2} has been included. As we can see, the pattern of convergence is the same in both cases, although the three-body force reduces the maximum of the correlation function by about 40\%. It should be noticed that including only the $1/2^+$ state, the correlation function goes to zero as the energy increases. On the other hand, at high energies the correlation function tends to one, this behavior is obtained after including a sufficient number of partial waves as we discuss below.

\begin{figure}[t!]
	\includegraphics[width=\columnwidth]{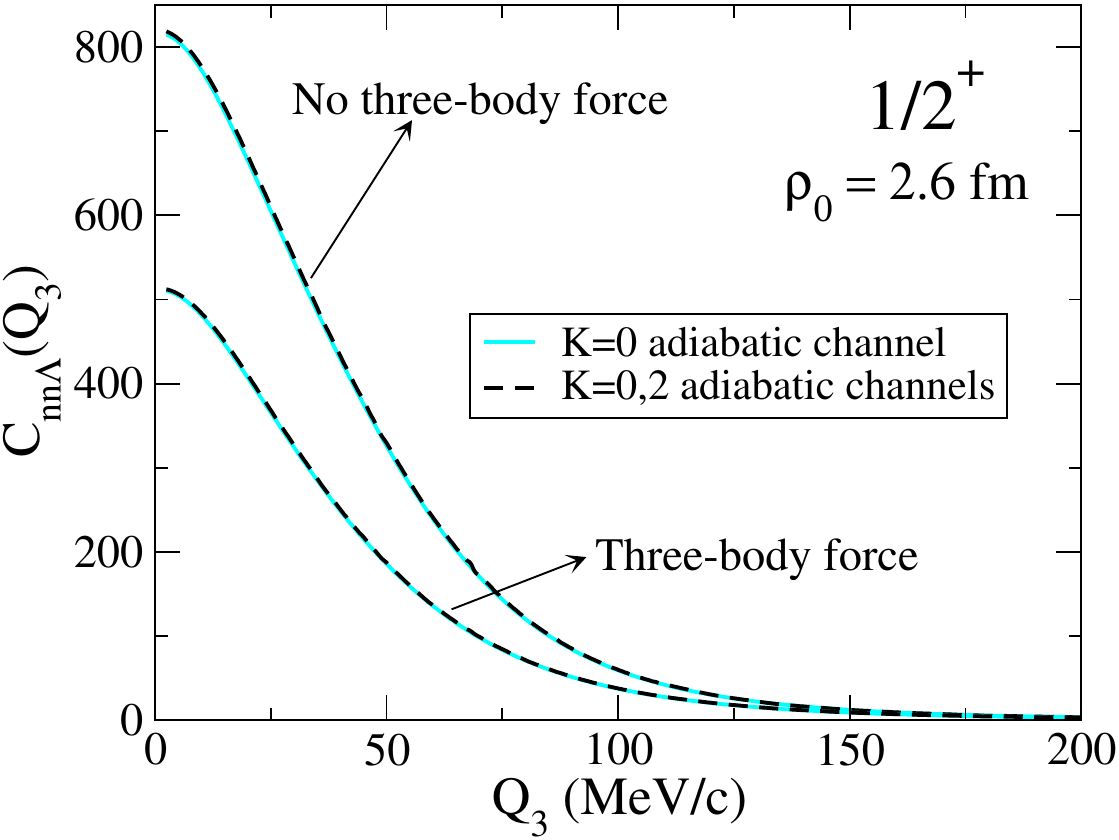}
	\caption{Converged $\nnL$ correlation function for the $1/2^+$ state including the lowest adiabatic potential ($K=0$), solid lines, and including the $K=0$ and $K=2$ adiabatic potentials, dashed lines. The source function, Eq.~(\ref{sour3b}), has been taken with $\rho_0=2.6$ fm. The $\nL$ interaction has been taken to be the NLO19 potential with $C=600$ MeV given  in Table~\ref{Tab2}. Results with and without three-body force are shown.}
	\label{fig:adia2}
\end{figure}

In order to verify that the truncation of the HA expansion to the first adiabatic
term is enough in the description of the correlation function, in Fig.~\ref{fig:adia2} we compare the converged results for the 1/2$^+$ state when only the first adiabatic potential is considered (solid curves) with the results when also the adiabatic channels with $K=2$ are included (dashed curves). As in Fig.~\ref{fig:nnLKmax}, the Gaussian representation of the NLO19 potential with $C=600$ MeV is used, and the source function is taken with $\rho_0=2.6$ fm. As it can be seen in the figure, the contribution of the $K=2$ channels is already extremely small, and therefore the $K\geq 2$ adiabatic channels can be safely disregarded.

\begin{figure}[t]
	\includegraphics[width=\columnwidth]{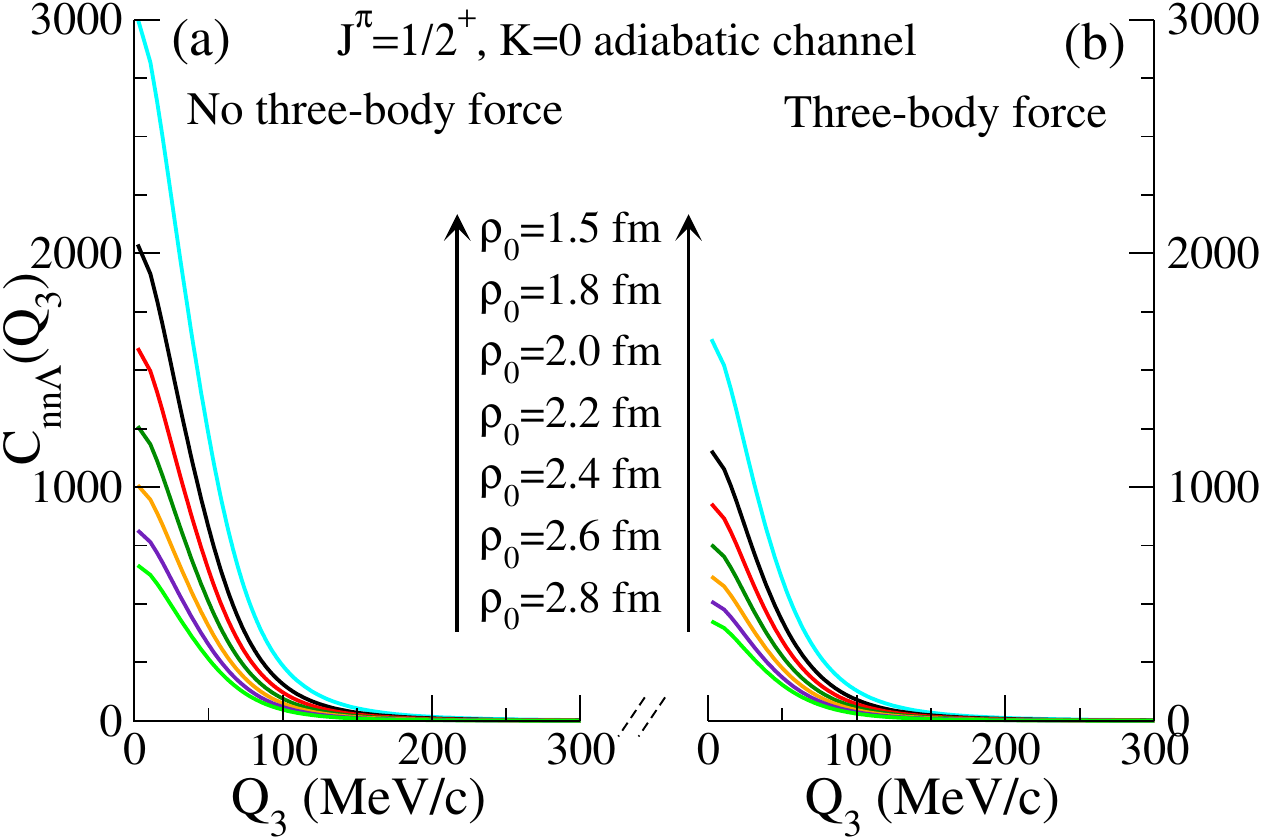}
	\caption{Converged $\nnL$ correlation function for the $1/2^+$ state including the lowest adiabatic potential ($K=0$) only. The $\nL$ interaction has been taken to be the NLO19 potential with $C=600$ MeV given 
 in Table~\ref{Tab2}. Panels (a) an (b) correspond to calculations without and with three-body force, respectively. The different curves are the calculations with
different values of $\rho_0$ used in the source function. As indicated by the arrow in the figure, the smaller $\rho_0$ the higher the maximum of the correlation function. }
	\label{fig:nnrho}
\end{figure}

Although the source radius of the three-body correlation function is related to the single particle source radius, $\rho_0 \approx 2 R_m$ \cite{pdtheory,pppth}, 
we would like to analyze the sensibility to variations of $\rho_0$. This is
done in Fig.~\ref{fig:nnrho}, where different values of $\rho_0$ have been used
to compute the converged correlation function for the 1/2$^+$ state. Again, Figs.~\ref{fig:nnrho}a and \ref{fig:nnrho}b show the results without and with three-body force, respectively. As we  can see, there is a big sensibility to this parameter. 

\begin{figure}[t]
	\includegraphics[width=\columnwidth]{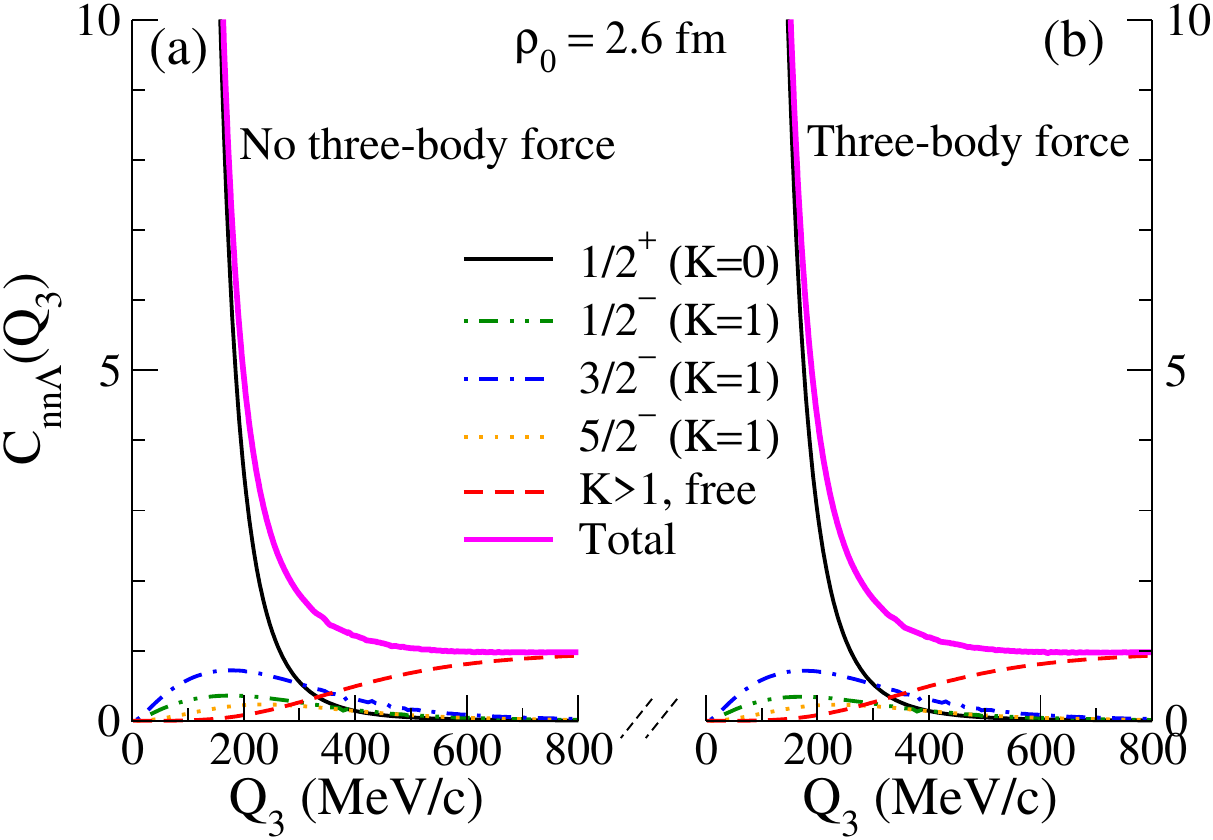}
	\caption{$\nnL$ correlation function for the $1/2^+$, $1/2^-$, $3/2^-$, and $5/2^-$ states including only the lowest adiabatic potential for each of them ($K=0$ and $K=1$ adiabatic potentials). The remaining, $K>1$, adiabatic channels are included as free (dashed red curve). The thick purple curve gives the total correlation function. The source function has been taken with $\rho_0=2.6$ fm. The $\nL$ interaction has been taken to be the NLO19 potential with $C=600$ MeV given in Table~\ref{Tab2}. Panels (a) an (b) correspond to calculations without and with three-body force, respectively.}
	\label{fig:nnLfull}
\end{figure}

Finally, in Fig.~\ref{fig:nnLfull} the complete correlation function is
calculated with the interaction active in the states with 
$J^{\pi}=1/2^+,1/2^-,3/2^-,5/2^-$ corresponding to the inclusion of the
interaction in $s$- and $p$-waves. Starting from $K>1$ the scattering wave
function is considered as free. From the figure we can observe that the state
$J^\pi=1/2^+$ is clearly dominant. The $p$-wave states are noticeably smaller, and starting from $K=2$ it is safe to consider the scattering state as non interacting. In fact, in order to make the $p$-wave curves visible we just show in the figure a zoom of the correlation function, such that only the function for $Q_3>200$ MeV/c is shown. As seen in Fig.~\ref{fig:adia2}, this $Q_3$ region corresponds to the very tail of the correlation function, where the effects of the three-body force are barely visible. The consequence is that Figs.~\ref{fig:nnLfull}a and \ref{fig:nnLfull}b, which show the results without and with three-body force, look pretty much the same, even if the three-body force drastically reduces the maximum of the correlation function, as seen in Figs.~\ref{fig:nnLKmax}, \ref{fig:adia2}, and \ref{fig:nnrho}. Moreover, the correlation function tends asymptotically to one as now we are including analytically all partial waves with $K>1$.

\section{The $\ppL$ system}
\label{sect5}

\subsection{Treatment of the Coulomb interaction}

Before entering into the details of the $\ppL$ correlation function it is important
to have in mind that the Coulomb repulsion between the two protons, 
\begin{equation}
V_\mathrm{Coul}(r)= \frac{e^2}{r},
\end{equation}
has to be taken into account, $r$ being the distance between the two protons. The presence of this long-range force
asymptotically couples the different partial waves of the three-body continuum wave 
function. The consequence is that the asymptotic part of the three-body wave functions can not be described analytically. This fact makes rather difficult the extraction of the $T$-matrix elements, which are necessary to provide the correct normalization of the wave function, as shown in Eq.(\ref{eq:asymu}) for the $\nnL$ case.

In Ref.~\cite{pppC} different ways to overcome this difficulty have been shown. In particular it was shown that, in the case of the correlation
function and due to the finite size of the source, it is possible to 
screen the Coulomb force. Accordingly the Coulomb potential is modified as
\begin{equation}
        V_\mathrm{sc}(r)=\frac{e^2}{r}\, e^{-(r/r_\mathrm{sc})^n} \,,
        \label{cscr}
\end{equation}
where $r_\mathrm{sc}$ is the screening radius and the parameter $n$ allows for a sufficient fast cut of the Coulomb potential (the value $n=4$ can be
safely used, see for example Ref.\cite{kievsky2010}). 

In Ref.~\cite{pppC} the effects of using a screened Coulomb potential to describe the $pp$ and $ppp$ correlation functions is discussed. Here we only show the main results of this study for the $\pp$ correlation function, which is given by Eq.(\ref{corr12}), with $r_0=R_M$, see Eq.(\ref{eq:r0}), and with 
$R_M$ the source radius and $\Psi_k(r)$ the $\pp$ scattering wave function.

To calculate the $\pp$ scattering wave function we consider 
the following Gaussian interaction plus the screened Coulomb potential
\begin{equation}
	V_{\pp}(r)=V_0 e^{-(r/r_0)^2} {\cal P}_0 + V_\mathrm{sc}(r) \ ,
\label{eq:gausspp}
\end{equation}
where the parameters of the Gaussian interaction are the same as the ones used in the $nn$ case, $V_0=-30.45\,$MeV and $r_0=1.815\,$fm,
(${\cal P}_0$ is a projector on spin $S=0$ state). With this parameterization, and in conjunction 
with the pure Coulomb interaction, this potential reproduces the experimental values of the $\pp$ scattering length and effective range,
as well as the $pp$ correlation function \cite{kievsky2004}. 
It can be considered a low-energy representation of the $\pp$ interaction as it
has been discussed in Ref.~\cite{tumino2023}. 

\begin{figure}[t]
	\includegraphics[width=\columnwidth]{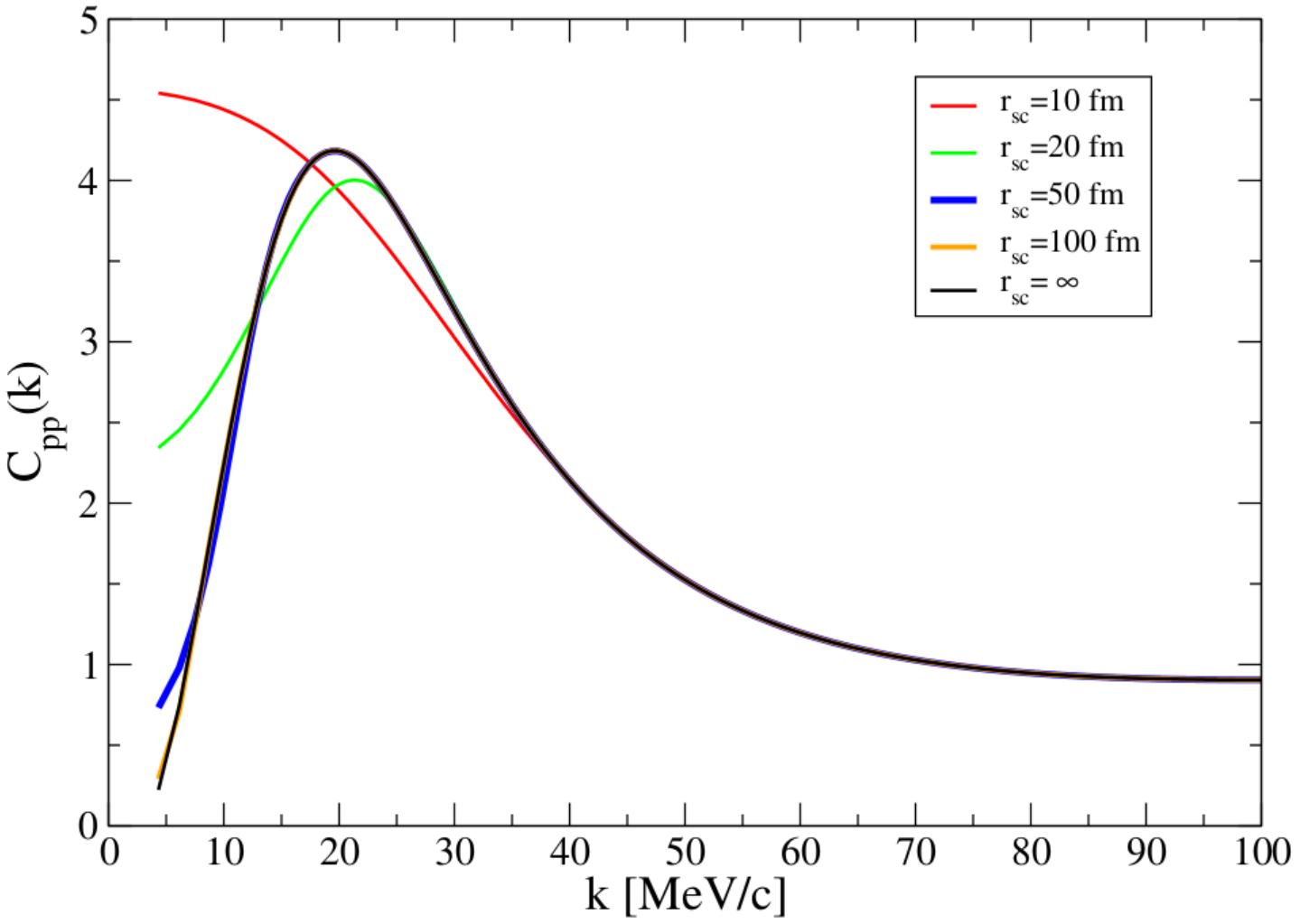}
        \caption{The $\pp$ correlation function for different screening radius. The case $r_{sc}=\infty$ corresponds to the use of the bare Coulomb force.}
        \label{fig:ppsc}
\end{figure}

In Fig.~\ref{fig:ppsc} we show the $\pp$
correlation function for a source radius of $R_M=1.249\,$fm and 
using different $r_\mathrm{sc}$ values. For two-body systems the asymptotic behavior of the
continuum wave functions is known also for charged particles, and the calculation using
the bare Coulomb force can therefore be performed. The result is the curve quoted in the figure with $r_\mathrm{sc}=\infty$. From the figure we can observe that, as the value of the screening radius
increases, the effects on the correlation function are appreciable at lower and
lower momenta, in such a way that the use of a screening radius of about 
$r_\mathrm{sc}=100$~fm 
produces results almost indistinguishable from those using the bare Coulomb force.
It should be noted that when using the screened Coulomb potential the effective proton-proton
interaction is short-range, and therefore the continuum wave function
tends asymptotically to a combination of Bessel functions. If the screening is not used,
the asymptotic behavior is given by a combination of Coulomb functions, known at the two-body level, but unknown for three-body systems.

\subsection{The correlation function}

After clarifying the treatment of the $\pp$ interaction with the screened Coulomb
potential we enter into the discussion of the $\ppL$ correlation function.
Since the screened Coulomb potential $V_\mathrm{sc}(r)$ is short-range, the equations
for the correlation function
used in the $\nnL$ case are applicable as well to the
$\ppL$ system with a screened potential. Therefore, as in Eq.{(\ref{cnnL})}, the $\ppL$ correlation function is given by 
\begin{eqnarray}
\lefteqn{
C_{\ppL}(Q)= \frac{1}{4}\frac{2^6}{Q^4\rho_0^6} 
\int \rho\, d\rho\, e^{-\frac{\rho^2}{\rho_0^2}} \times }
\label{cppL} \\ &&
	 \left(\sum_J	(2J+1)\left|\frac{u_{n_0}^J}{\sqrt{Q\rho}}\right|^2
	+  \sum_{K>1} J^2_{K+2}(Q\rho) N_{ST}(K)
	\right) ,
\nonumber
\end{eqnarray}
where the sum over $J$ includes the states $J^\pi=1/2^+,1/2^-,3/2^-,5/2^-$ 
with $u_{n_0}^J$ the corresponding wave function calculated using the lowest adiabatic
channel. 

To study this correlation function, we consider the different $\pL$ 
interactions defined in Section~\ref{sect2} and we use the Gaussian representation of the
$\pp$ interaction as given in Eq.~(\ref{eq:gausspp}). In this way we can study the
sensitivity of the $\ppL$ correlation function to the different
descriptions of the $\pL$ potential. 

\begin{figure}[t]
        \includegraphics[width=\columnwidth]{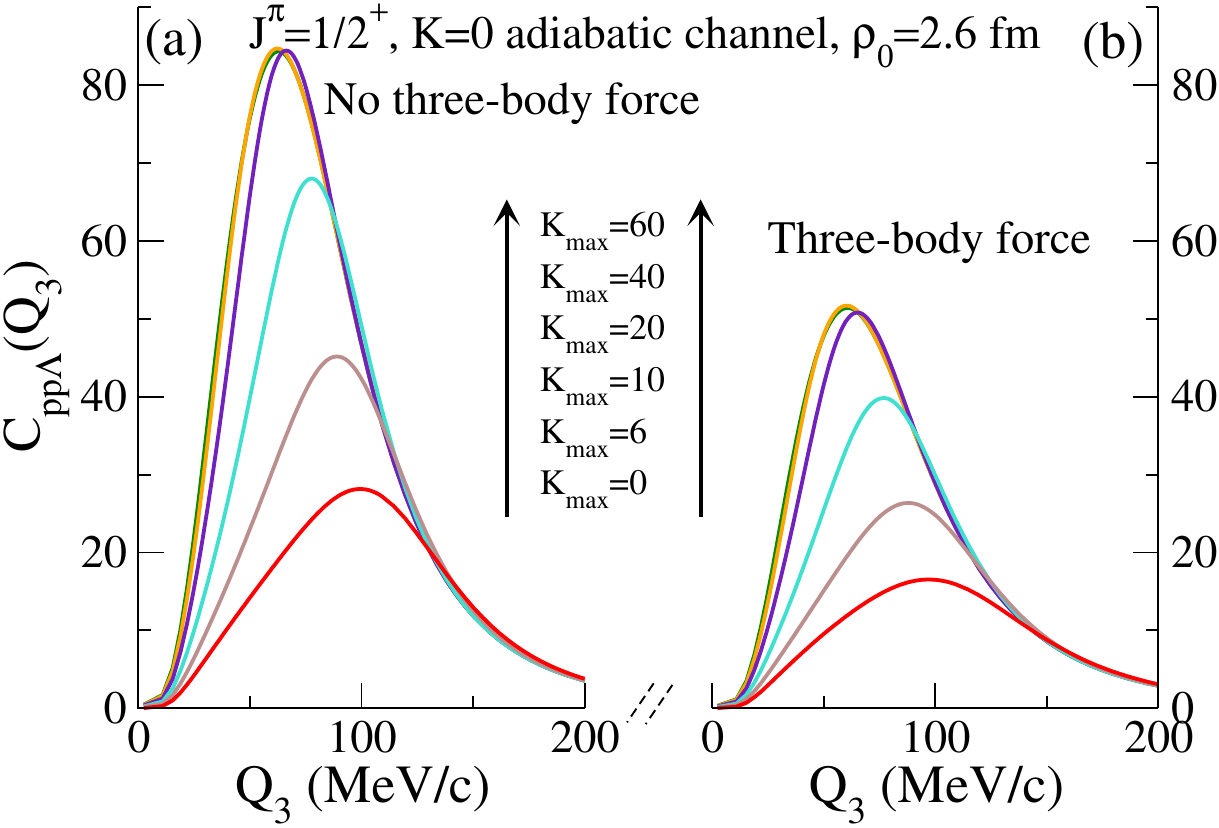}
        \caption{Same as Fig.2 but for the $\ppL$ correlation function.}
        \label{fig:ppLKmax}
\end{figure}

We start the study of the $\ppL$ correlation function looking at the
$J^\pi=1/2^+$ state, which gives the larger contribution to the
observable. We first consider calculations including only the lowest adiabatic
potential, asymptotically associated to a single HH with $K=0$.  To investigate
the convergence with the number of HH used to describe 
the lowest adiabatic channel, we show the correlation function in Fig.~\ref{fig:ppLKmax} for different values of the maximum value of the grand-angular quantum number, $K_\mathrm{max}$,
used in the expansion (\ref{bastr}). The NLO19 $\pL$ interaction with cutoff $C=600$ MeV,
see Table~\ref{Tab2}, has been used. The source function has been taken with
$\rho_0=2.6$ fm. As we can see in the figure, the convergence of the $\ppL$ correlation
is achieved for $K_\mathrm{max}$ values of about 60, clearly smaller than in the
$\nnL$ case, where $K_\mathrm{max}\approx 500$ was needed. Figs.~\ref{fig:ppLKmax}a and \ref{fig:ppLKmax}b show the results without and with three-body force,
respectively.

\begin{figure}[t]
        \includegraphics[width=\columnwidth]{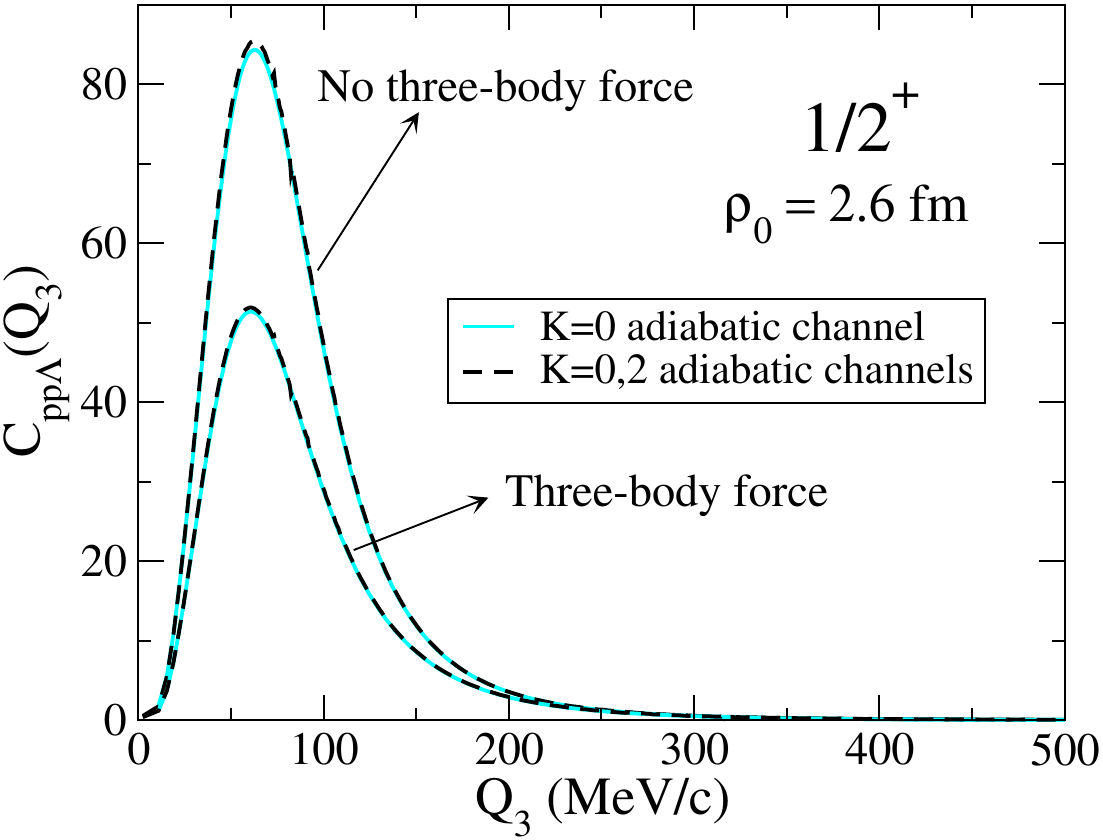}
\caption{Same as Fig.3 for the $\ppL$ correlation function.}
        \label{fig:ppadia2}
\end{figure}

 In order to check that the inclusion of the first adiabatic channel only in the adiabatic expansion, Eq.~(\ref{3bdad}), is sufficiently accurate, we in Fig.~\ref{fig:ppadia2} compare
 these results (solid curves) with the ones obtained when the contribution from the adiabatic channels associated to $K=2$ are also included (dashed curves). We observe
that, as in the $\nnL$ case, inclusion of the next adiabatic terms give an almost negligible contribution.

\begin{figure}[t]
        \includegraphics[width=\columnwidth]{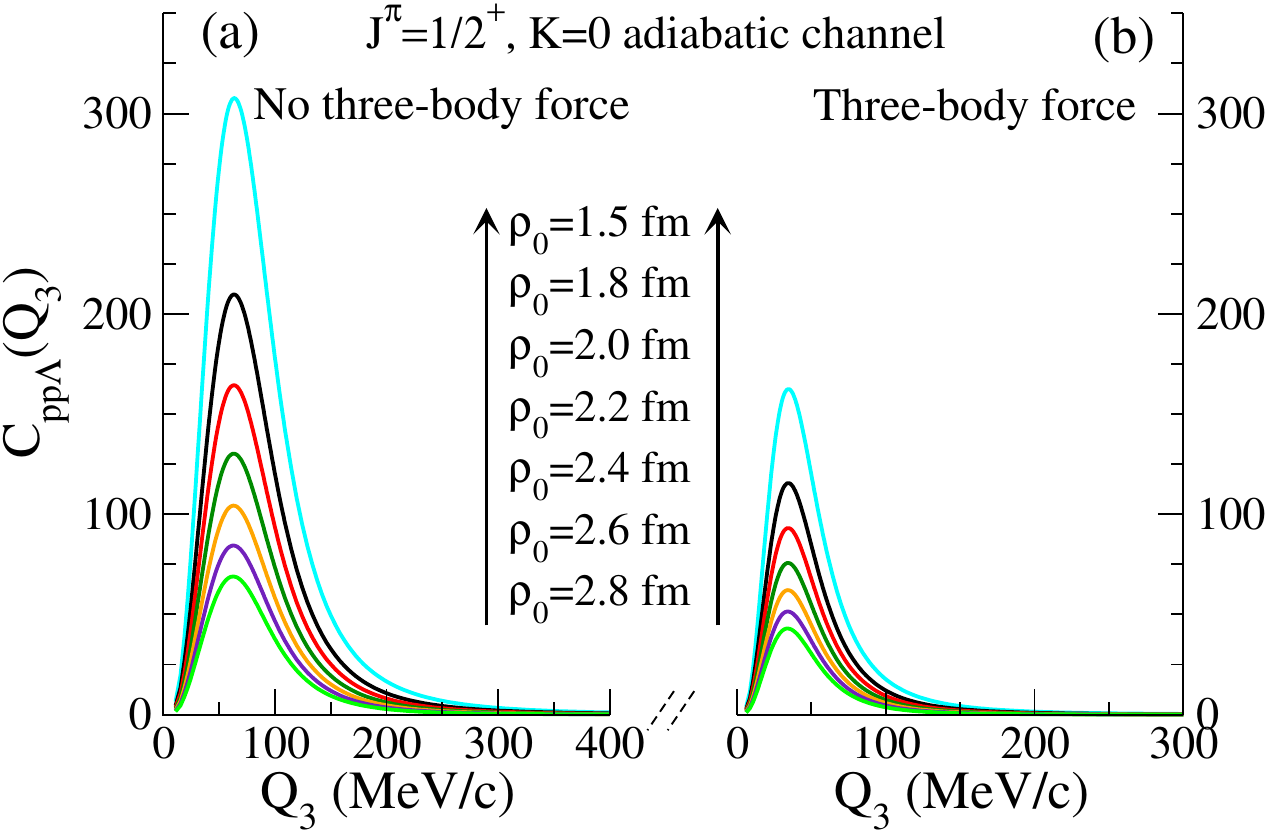}
\caption{Same as Fig.4 for the $\ppL$ correlation function}
        \label{fig:pprho}
\end{figure}

\begin{figure}[t]
        \includegraphics[width=\columnwidth]{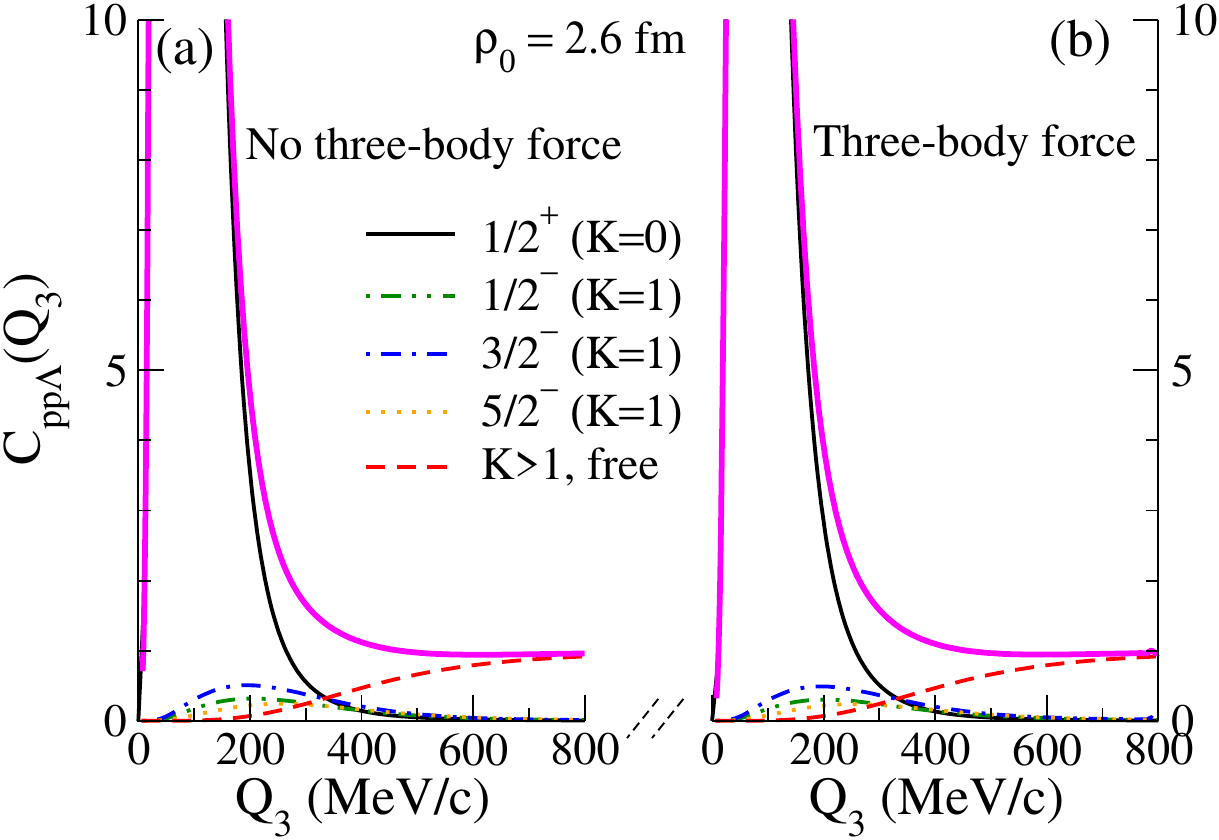}
\caption{Same as Fig.5 but for the $\ppL$ correlation function}
        \label{fig:ppLfull}
\end{figure}

In Fig.~\ref{fig:pprho} the variation of the correlation function is studied for
different source radii. As in the $\nnL$ system, we find a large dependence on the $\rho_0$ value. The three-body force reduces significantly the peak of the correlation function.

In Fig.~\ref{fig:ppLfull} we study the contributions of the $J^\pi=1/2^+,1/2^-,3/2^-,5/2^-$ states. The main result is also similar to
the $\nnL$ case, that is, the negative parity states give a much
smaller contribution to the correlation function whereas the huge peak at low momentum
is by far dominated by the $J^\pi=1/2^+$ state.

\begin{figure*}[t]
\includegraphics[width=16cm]{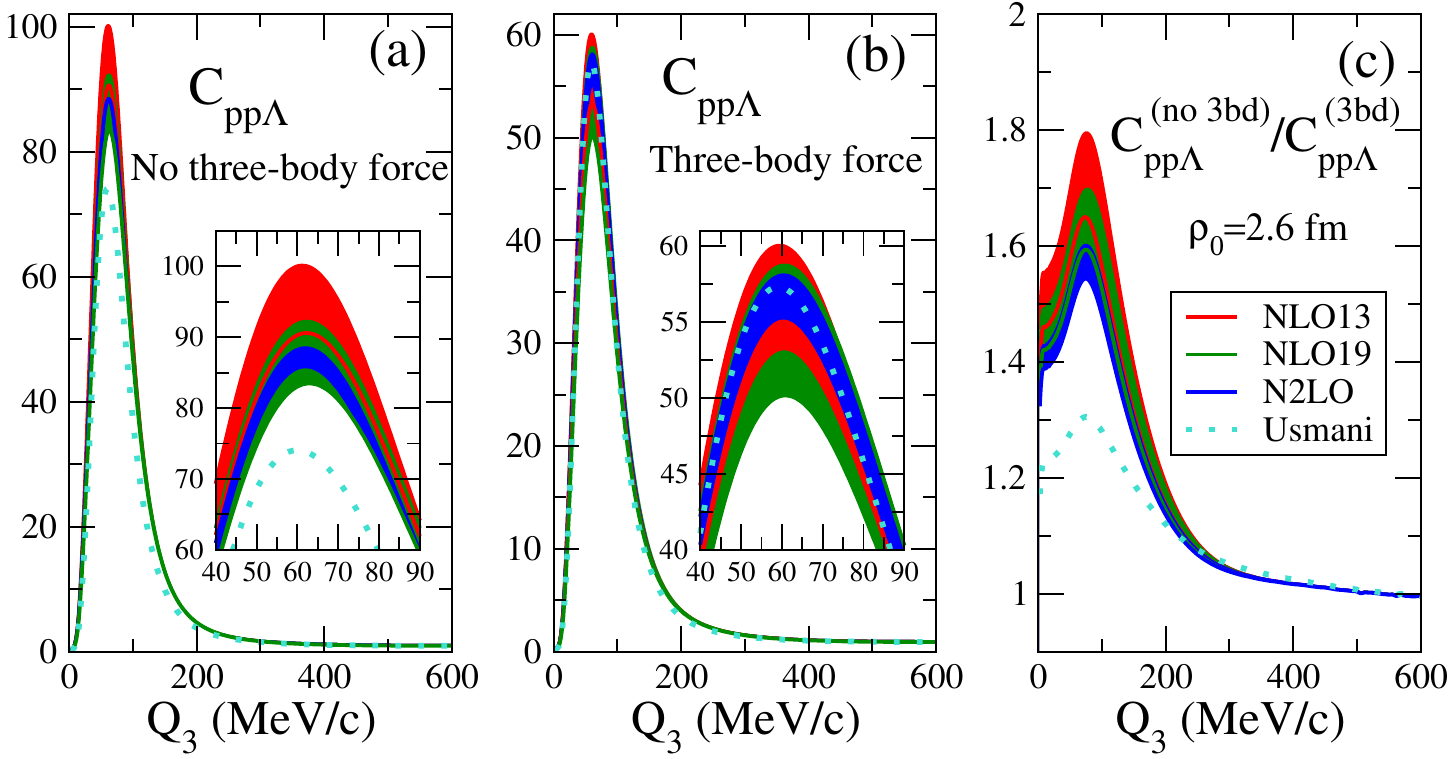}
	\caption{Total $\ppL$ correlation function calculated with the indicated potential models. (a) Only two-body forces are used. (b) Three-body force has been included. (c) Ratio between the two curves for each potential model.}
	\label{fig:ppl}
\end{figure*}

Finally, in Fig.~\ref{fig:ppl} we show the results for the total correlation function using all the 13 Gaussian two-body forces given in Table~\ref{Tab2}. The results with the NLO13, NLO19, and N2LO potentials are shown by the solid, dot-dashed, and dashed curves, respectively. Figs.~\ref{fig:ppl}a and\ref{fig:ppl}b are the results without and with three-body force, and the inset shows a zoom of the peak region. In Fig.~\ref{fig:ppl}c we show the ratio between the correlation functions without and with three-body force. For completeness we show as well the results obtained when using the Bodmer-Usmani two-body potential, Eq.(\ref{bodus}), (dotted curve). In this case the three-body force is of the Wigner type (see Ref.~\cite{bodmer1})
\begin{equation}
    W(r_{13},r_{23})=W_3^0 \, T_\pi^2(r_{13})\, T_\pi^2(r_{23}) \ ,
\end{equation}
with the strength $W_3^0=0.025\,$MeV fixed to reproduce the hypertriton binding energy and $T_\pi(r)$ being the pion-exchange tensor potential \cite{bodmer1,bodmer}.

The main result is that the correlation
functions calculated with the different potentials almost overlap with a
small sensitivity to the interaction in the peak. The Gaussian potentials show a spread around $15\%$ on the peak mostly due to the spread of the curves based on the NLO13 scattering parameters. The results using the NLO19 and N2LO scattering parameters are much closer to each other whereas the Bodmer-Usmani potential is slightly lower. When the three-body interaction is included we can observe a consistent reduction of the observable (around $40\%$), together with a reduction of the spread, which is now of about $10\%$. The NLO13 curves are still on the upper part and close to those of the N2LO interactions as well as to the Bodmer-Usmani curve. The NLO19 curves are on lower part of the peak. The last panel shows, for the different potentials,
the ratios between the correlation functions without and with three-body force, which go asymptotically to 1. As we can see in the figure, the result obtained
with the Bodmer-Usmani potential is slightly different because the maximum of the correlation function without three-body force, Fig.~\ref{fig:ppl}a, is clearly lower than in the other cases, although this difference disappears when the
three-body force is included. In the figure the calculations have been done using the source size of $\rho_0=2.6\,$fm.

\section{Comparison to experimental data}
\label{sect6}
The experimental correlation function measured by ALICE includes genuine $\ppL$ contributions as well as feed-down correlations and misidentified particles. The dominant feed-down contributions to the $\ppL$ triplets in the data arise from the decay of $\Sigma^0$ and $\Xi$ resonances into $\Lambda$ hyperons, whereas all the remaining contributions are negligible and considered as flat correlations. 

Similarly to what described in Sec.~\ref{sec:Results2B}, the correlation function is decomposed as
\begin{eqnarray}
 C_{\ppL}(Q_3) &=& \lambda_{\ppL} C_{\ppL}^\mathrm{th}(Q_3) 
               + \lambda_{\ppL_{\Sigma^0}} C_{\ppL_{\Sigma^0}}(Q_3) \nonumber \\ 
               && + \lambda_{\ppL_{\Xi}} C_{\ppL_{\Xi}}(Q_3) 
               + \lambda_\mathrm{flat}, 
               \label{eq:ppLdec}    
\end{eqnarray}
where the $\lambda_i$ parameters are the weights of each contribution, $C_{\ppL}^\mathrm{th}(Q_3)$ is the genuine computed correlation function and $C_{\ppL_\mathrm{X}}(Q_3)$ are the residual feed-down correlations from  the decay of the resonance X=\{$\Sigma^0, \Xi$\}. 

In the two-body $\pL$ correlation function a non-flat feed-down contribution was considered because the precision of the experimental data provides sensitivity to the momentum dependence of such an effect. In the $\ppL$ measurement, a flat feed-down contributions can be assumed. This allows us to rewrite Eq.~\eqref{eq:ppLdec} as
\begin{equation}
  C_{\ppL}(Q_3) = \lambda_{\ppL} C_{\ppL}^\mathrm{th}(Q_3) + 1 - \lambda_{\ppL} \ .
\label{eq:3Bcorrection}
\end{equation}

The genuine contribution from $\ppL$ triplets is weighted by the factor $\lambda_{\ppL}$ = 0.405~\cite{femtoppp}. Non-femtoscopic background correlations in the measurement are absent, as shown in Ref.~\cite{femtoppp}, and the effect of the momentum resolution is found to be negligible. 

\begin{figure}[t]
 	\includegraphics[width=\columnwidth]{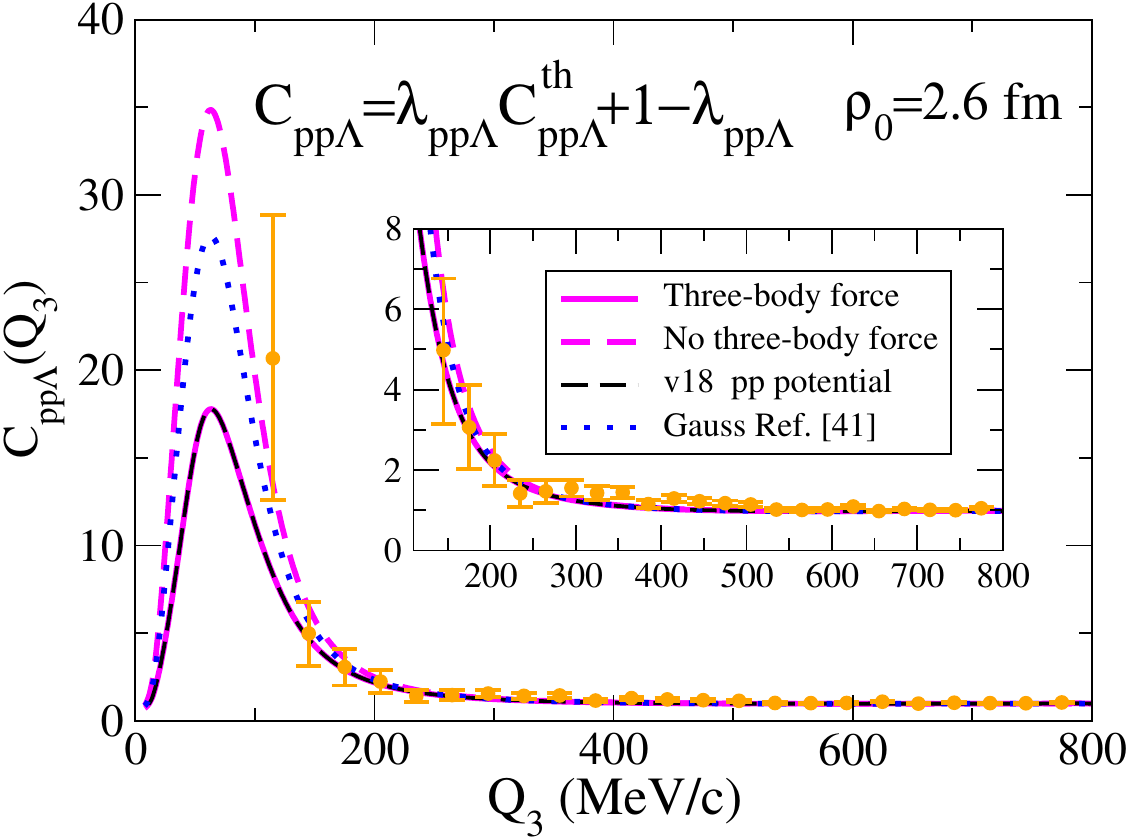}
	\caption{Comparison with the experimental data. For the $pp$ interaction Eq.(65) has been used whereas for the $\pL$ interaction the Gaussian representation of the NLO19 model with $C=600$ MeV has been considered with (pink solid line) and without (pink dashed line) including the three-body force. The black dashed line is the result using the Argonne $v_{18}$ $pp$ interaction and the three-body force. The result using the Gaussian potential obtained from the scattering parameters of Ref.~\cite{mihaylov2024} is also shown by the blue dotted line.}
	\label{fig:pplRm}
\end{figure}

The resulting correlation functions $C_{pp\Lambda}(Q_3)$ are shown in Fig.~\ref{fig:pplRm}. The figure shows the total $\ppL$ correlation function calculated using the Gaussian representation of the $pp$ interaction with Coulomb screening as given in Eq.(\ref{cscr}), and the NLO19 $\pL$ potential model with $C=600\,$ MeV. The results without (dashed pink line) and with (solid pink line) three-body force are shown. 

For completeness, we also show the results using the more realistic Argonne $v18$ $pp$ interaction (still with Coulomb screening). The result is shown by the black dashed curve which perfectly overlaps with the curve obtained with the simple $pp$ Gaussian potential. 
Finally, we have calculated the $pp\Lambda$ correlation function using the $p\Lambda$ scattering parameters from Ref. \cite{mihaylov2024} to construct the two-body Gaussian potentials in the $S=0$ and $S=1$ channels. The result is shown in Fig. \ref{fig:pplRm} with the blue dotted line. As discussed in Section \ref{sec:Results2B}, the hypertriton binding energy is close to the experimental value leaving room for a tiny repulsive three-body force. In this work we do not include the three-body force for this model because it would require a deeper study of its impact on the heavier hypernuclei measurements and eventually to introduce more complicated form of the potential.
Being less attractive, the peak described by this potential is slightly lower than the peak described by the Gaussian NLO19 model.

From the figure we can also see that much of the data are in the region in which the correlation function is close to one. However the data show the region in which the peak starts to appear (looking from high to lower values of $Q_3$) and there is one data point in the region of interest. This figure confirms the possibility of studying the $\pL$ and $\ppL$ interactions using the correlation function and indicates the best region in which experimental data and theoretical calculations can be compared.

\section{Summary and conclusions}
\label{sect7}

The present study describes the different steps followed to compute the $\ppL$ correlation function. The
correlation function is a convolution of the source function with the square of the scattering wave function. Accordingly part of the discussion has been focused in the description of the $\ppL$ scattering wave function.
To this aim we have decomposed the wave function in partial waves having well defined values of $J^\pi$. The lowest partial waves, which are affected by the interaction were described using the HA basis method. Partial waves with $J^\pi=3/2^+$ and higher, for positive parities, and with $J^\pi=7/2^-$ and higher, for negative parities have been considered as free. In the description of the lowest partial waves different interactions have been used and their impact on the correlation function has been evaluated. In particular low-energy representations of the $N\Lambda$ interaction have been used in the form of Gaussian potentials. They have been constructed to reproduce the scattering lengths and effective ranges in the two spin states as determined by potentials models based on chiral-EFT fitted to the $\pL$ scattering data. When these potentials were used in the three-body system a repulsive three-body force has been included. This term is needed in order to reproduce the hypertriton binding energy as well as those of the $^4_{\Lambda}$H and $^4_{\Lambda}$He hypernuclei.

Following the studies on the $ppp$ correlation function~\cite{pppth,pppC}, the Coulomb potential has been screened. In the description of processes in which the particles interact asymptotically with the Coulomb potential, a simple screening treatment is not enough~\cite{deltuva2008}. However in the present case, the finite size of the source allows for this kind of simplification of the asymptotic configuration. 

We can summarize the main results of the present work as follows. In the case of the two-body $\pL$ correlation function, the low-energy Gaussian interactions produce results within a narrow band. When this band is corrected from the residual induced correlations (see Section~\ref{sec:Results2B}), it can be compared to the experimental data. This comparison evidences deviations larger than 6.5$\sigma$ between the theoretical and experimental correlation functions in the low relative momentum region $k < 120$ MeV/c, motivating further studies of the low energy $p\Lambda$ interaction using femtoscopy measurements. Improvements in this direction have been recently done in~\cite{mihaylov2024} by modelling the $p\Lambda$ interaction from a combined analysis of the scattering and femtoscopy data. 
We stress that in the present study we did not include the $\Lambda-\Sigma$ mixing, a more detailed potential including both channels is postponed to a forthcoming study. 
The $\ppL$ correlation function shows a huge, low-energy peak, essentially produced by the $1/2^+$ partial wave. This peak dominates the correlation function at low energies and it is not affected by the other partial waves. On the other hand, the peak is strongly affected by the three-body force and by the size of the source. The peak shows some dispersion, of about $15\%$, when the different two-body potentials are used. This dispersion reduces below $10\%$ when the three-body force, fixed to describe the hypertriton binding energy, is included. To some extent this fact indicates that the correlation function, in the low-energy regime, is correlated to the hypertriton binding energy.

The comparison to the experimental data is discussed in Section VI and, as in the two-body case, corrections from feed-down correlations and misidentified particles are necessary. We can observe that the available data suggest the formation of the low-energy peak, although experimental points below $100$ MeV/c are absent, without covering the main region where the peak is located. The particular structure of the $\ppL$ correlation function at low energies, strongly dominated by one partial 
wave, is well suited for detailed experimental as well theoretical studies from which important conclusions can be extracted regarding the $\pL$ and $\ppL$ interactions. To this respect a deeply study of the $\ppL$ correlation function as well as more data in the low energy region would be very welcome.

\acknowledgments   
This work has been partially supported by: Grant PID2022-136992NB-I00 funded by MCIN/AEI/10.13039/501100011033 and, as appropriate, by “ERDF A way of making Europe”, by the “European Union” or by the “European Union NextGenerationEU/PRTR”; the Deutsche Forschungsgemeinschaft through Grant SFB 1258 “Neutrinos and Dark Matter in Astro- and Particle Physics”; the Deutsche Forschungsgemeinschaft (DFG, German Research Foundation) under Germany's Excellence Strategy – EXC 2094 – 390783311.

\bibliographystyle{apsrev4-1} 
\bibliography{refs} 

\end{document}